\documentclass[twocolumn,showpacs,amsmath,amssymb,superscriptaddress]{revtex4}
\usepackage{graphicx,amsmath,color}

\newcommand{\unit}[1]{\,{\rm #1}}
\newcommand{\sub}[1]{_{\rm #1}}
\newcommand{\subup}[1]{^{\rm #1}}

\newcommand{\vctr}[1]{{\bf {#1}}}

\newcommand{\ann}[1]{#1^{\phantom{\dagger}}}
\newcommand{\cre}[1]{#1^{\dagger}}

\begin{document}

\title{Singlet-triplet splitting in double quantum dots due to spin orbit and hyperfine interactions}

\author{Dimitrije Stepanenko}
\affiliation{Department of Physics, University of Basel,
             Klingelbergstrasse 82, CH-4056 Basel, Switzerland}
\author{Mark Rudner}
\affiliation{Department of Physics, Harvard University, 17 Oxford St., 5 Cambridge, MA 02138, USA}
\author{Bertrand I.~Halperin}
\affiliation{Department of Physics, Harvard University, 17 Oxford St., 5 Cambridge, MA 02138, USA}
\author{Daniel Loss}
\affiliation{Department of Physics, University of Basel,
             Klingelbergstrasse 82, CH-4056 Basel, Switzerland}

\date{\today}


\begin{abstract}

We analyze the low-energy spectrum of a two-electron double quantum
dot under a potential bias in the presence of an external magnetic
field.  We focus on the regime of spin blockade, taking into account
the spin orbit interaction and hyperfine coupling of electron and
nuclear spins.  Starting from a model for two interacting electrons in
a double dot, we derive a perturbative, effective two-level
Hamiltonian in the vicinity of an avoided crossing between singlet and
triplet levels, which are coupled by the spin-orbit and hyperfine
interactions.  We evaluate the level splitting at the
anticrossing, and show that it depends on a variety of parameters
including the spin orbit coupling strength, the orientation of the
external magnetic field relative to an internal spin-orbit axis, the
potential detuning of the dots, and the difference between hyperfine
fields in the two dots.  We provide a formula for the splitting in
terms of the spin orbit length, the hyperfine fields in the two dots,
and the double dot parameters such as tunnel coupling and Coulomb
energy.  This formula should prove useful for extracting spin orbit
parameters from transport or charge sensing experiments in such
systems. We identify a parameter regime where the spin orbit and
hyperfine terms can become of comparable strength, and discuss how
this regime might be reached.  
\end{abstract}


\maketitle

\section{Introduction}

Electron spins in gated quantum dots have been extensively studied for
their possible use in quantum information processing
\cite{LD98,HKP+07,ZRC+10}.  In this context the main interest lies in the
study of coherent quantum evolution of electron spins in a network of
coupled quantum dots in the presence of external magnetic fields.  A
double quantum dot (DQD) populated by two electrons is the smallest
such network in which all of the steps necessary for quantum
computation can be demonstrated.  In addition, a DQD can host encoded
two-spin qubits which require less resources for control than the
single-electron spins in quantum dots.  In DQDs, the spins can be
manipulated exclusively by electric fields in the presence of a constant
magnetic field, taking advantage of the spin orbit and/or nuclear
hyperfine interactions
\cite{HKP+07,B01,L02,WL02,SB04,GBL06,KBT+06,NKN+07,BRM+09,FBM+09,NFB+10}.  

High precision requirements for the control of spin qubits have
prompted the detailed study of the interactions of electron spins in
quantum dots.  The DQDs give experimental access to the coherent spin
dynamics.  Studies of transport through DQDs in the spin blockade
regime \cite{OAT+02} have been particularly useful for probing the
electron and nuclear spin dynamics.  In this regime, charge transfer
between the two dots of a DQD can take place only when the electrons
form a singlet state with total spin zero.  This allows weak spin
non-conserving interactions to be studied via charge sensing
\cite{JPM+05} or by charge transport measurements
\cite{OAT+02,TKK+06}, even in the presence of  much stronger
spin-conserving interactions.  The most important spin non-conserving
interactions are  the spin orbit interaction and the  hyperfine
interaction between the electron spins  and a collection of nuclear
spins inside the DQD \cite{NFT+10}. 

In this work, we investigate the hyperfine and spin-orbit mediated
coupling between electronic singlet and triplet spin states of a DQD
in the spin-blockade regime.  We show that the spin-orbit and
hyperfine contributions to this splitting can be tuned
by a number of parameters.  We
derive an explicit formula that gives this splitting as a function of
a homogeneous external magnetic field and the detuning between the
ground-state energies of the dots in
a DQD.  These parameters can be varied in an experiment.  In addition,
the splitting depends on the spin-orbit coupling interaction and the
inhomogeneous nuclear Overhauser field, as well as on the dot
parameters such as the hopping amplitude between the dots in a DQD,
Coulomb repulsion between the electrons, and the direct exchange
interaction.  Further, we describe how the dependence of the
singlet-triplet splitting on these parameters might be used to extract
the intrinsic strengths of the spin orbit and hyperfine couplings from
charge sensing measurements in which the DQD is swept through a
singlet-triplet level crossing in the presence of spin-orbit
interaction and a fluctuating nuclear field.  

Recently, it was predicted that the angular momentum transferred
between electron and nuclear spins in both dc transport \cite{RL10}
and Landau-Zener-type gate sweep experiments \cite{RNL+10,BR11} can
show extreme sensitivity to the ratio of spin-orbit and hyperfine
couplings.  Our result gives an explicit dependence of this ratio on
the detuning and external magnetic field, thus showing how all regimes
can potentially be reached.  

The paper is organized as follows.  In Sec.~\ref{sec:model} we
introduce a model of a  DQD,  and describe its energy levels as a
function of detuning. In Sec.~\ref{sec:soi} we find the matrix
elements of the spin orbit interaction in the space of relevant
low-energy states.  In Sec.~\ref{sec:eh} we study the orbital and spin
structure of the singlet and triplet states which nominally intersect
for particular combinations of the DQD potential detuning and external
magnetic field.  In Sec.~\ref{sec:heff}, we define an effective
Hamiltonian which describes the action of the spin-orbit and hyperfine
couplings in the corresponding two-level subspace.  Then, in
Sec.~\ref{sec:splitting} we study the dependence of the resulting
singlet-triplet splitting on external parameters and show how this
dependence can be used to extract the spin orbit interaction strength
and the size of Overhauser field fluctuations from charge sensing
measurements.  In Sec.~\ref{sec:switching} we discuss how the DQD can
be tuned between the regimes of spin-orbit-dominated splitting and the
hyperfine-dominated splitting.  Finally, we summarize our results in
Sec.~\ref{sec:conclusions}.

\section{Model Hamiltonian for Double Quantum Dots}
\label{sec:model}

\begin{figure}[t]
\begin{center}
\includegraphics[width=8.5cm]{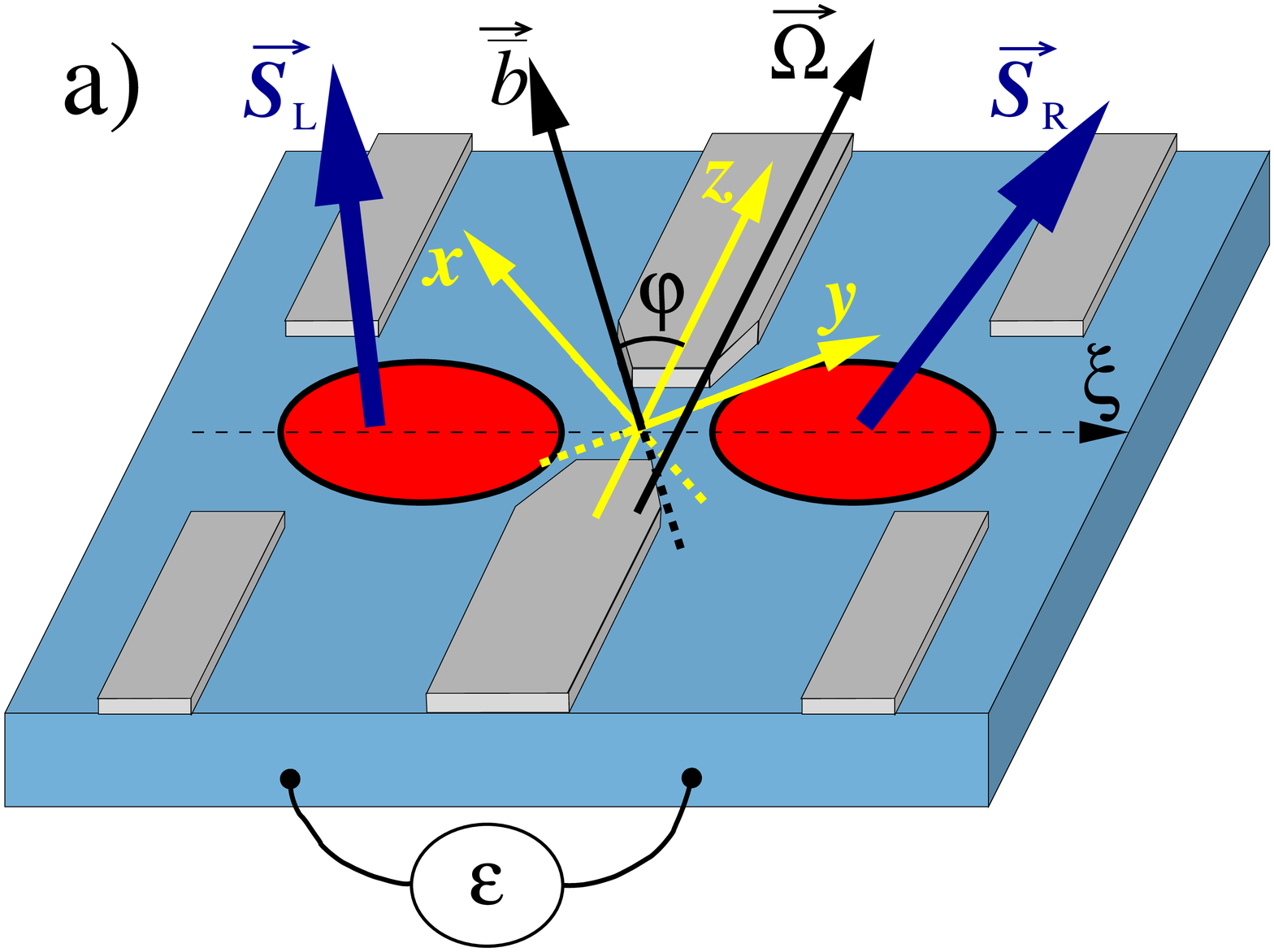}
\vskip 0.4cm
\includegraphics[width=8.5cm]{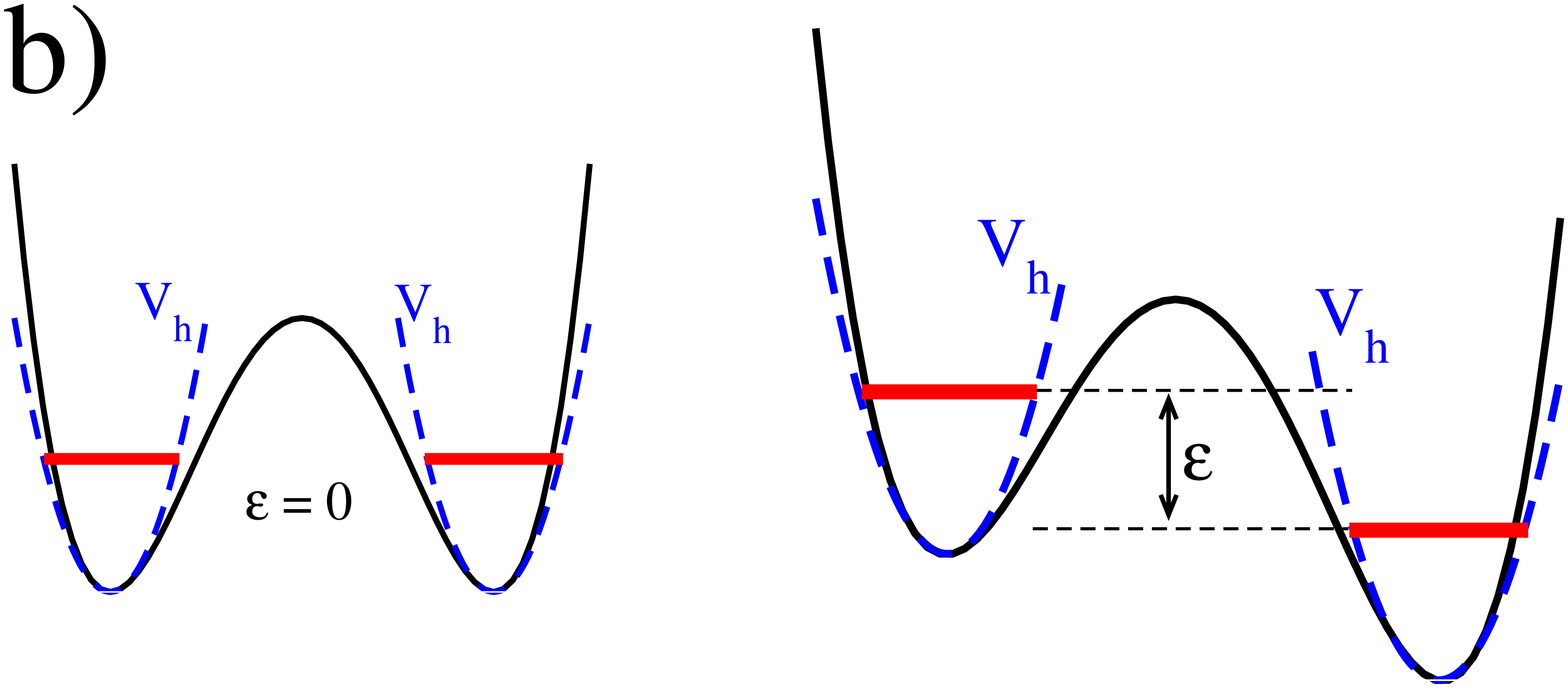}
\caption{\label{fig:axes} a) double quantum dot model
  and the coordinate system.  ${\bf S}_{L} $ and ${\bf S}_{R} $ denote
  the spin-$1/2$ of the electron in the right and in the left quantum
  dot, respectively.  The dots lie in the $\xi z$-plane and are
  tunnel-coupled  along the $\xi$-direction (perpendicular to the
  $z$-axis). They can be detuned by the externally applied voltage
  $\varepsilon/e$.  The spin orbit field $\boldsymbol{\Omega}$ points
  along the $z$-axis, defining the first quantization axis for the
  triplet states $|T_{z,\pm}\rangle$ and $|T_{z0}\rangle$  (see main
  text).  The effective magnetic sum-field $\bar{\vctr{b}}$ defines
  the second  quantization axis for the triplet states
  $|T_{\pm}\rangle$ and $|T_0\rangle$.  We choose the mutually
  orthogonal axes $x,y,z$ so that  $\bar{\vctr{b}}$ lies in the
  $xz$-plane.  b) effect of
  detuning on the quantum dot levels.  At zero detuning
  $\varepsilon=0$, an electron has the same energy on the left and
  right quantum dot.  For nonzero detuning, the energy of an electron
  on the left dot is $\varepsilon$ higher than the energy of an
  electron on the right dot.} 
\end{center}
\end{figure}

In a DQD, electrons are confined near the minima of a double-well
potential $V\sub{DQD}$, created by electrical gating of a
two-dimensional electron gas (2DEG) in, e.g.~GaAs, see
Fig.~\ref{fig:axes}.  For the case of a deep potential, we treat the two
local minima of the double-well as isolated harmonic
wells with ground state wave functions $\varphi_{1,2}$.  In order to
define an orthonormal basis of single particle states for building-up
the two-electron states of the DQD, we form the Wannier orbitals
$\Phi_L$ and $\Phi_R$, centered in the left and right dots,
respectively \cite{BLD99}:
\begin{equation}
\label{eq:wannierlr}
\Phi\sub{L,R} = \frac{1}{\sqrt{1-2sg+g^2}}\left( \varphi\sub{1,2} - g
\varphi\sub{2,1}\right),
\end{equation}
where $s=\langle \varphi\sub{1}|\varphi\sub{2}\rangle =  \exp[- \left(
  a / a\sub{B} \right) ^2]$ is the overlap of the harmonic oscillator
ground state wave functions of the two wells,
$a\sub{B}= \sqrt{\hbar/m\omega_0}$ is the Bohr radius of a single
quantum dot, $\hbar \omega_0$ is the single-particle level spacing,
and $2a=l$ is the interdot distance.  The mixing factor of
the Wannier states is $g= ({1-\sqrt{1-s^2}})/{s}$.

The two electrons in the DQD are coupled by the Coulomb interaction,  
\begin{equation}
\label{eq:coulombinteraction}
C=\frac{1}{4\pi \kappa} \frac{e^2}{|\vctr{r}_1 - \vctr{r}_2 |},  
\end{equation}
where $\vctr{r}_1$ ($\vctr{r}_2$) is the position of electron 1 (2)
and $\kappa$ is the dielectric constant of the host material.  In this
work we consider the regime where the single-particle level spacing is
the largest energy scale,  in particular $\hbar \omega_0 \gg e^2/(4\pi
\kappa a)$.  In this case, and assuming that the hyperfine and
spin-orbit interactions are also weak, single-particle orbital
excitations can be neglected.  Therefore, the relevant part of the
two-electron Hilbert space is approximately spanned by Slater
determinants involving the Wannier orbitals $\Phi\sub{L,R}$.
Including spin, and using second quantization notation where $\cre{c}_L$
($\cre{c}_R$) creates an electron in the Wannier state $\Phi_L$
($\Phi_R$), we define the two-electron basis states:
\begin{align}  
\label{eq:20singletdef}
|(2,0) S \rangle &= \cre{c}_{{\rm L} \uparrow} \cre{c}_{{\rm L}
  \downarrow} | 0 \rangle,
\\
\label{eq:02singletdef}
|(0,2) S \rangle &= \cre{c}_{{\rm R} \uparrow} \cre{c}_{{\rm R}
  \downarrow} | 0 \rangle,
\\
\label{eq:11singletdef}
|(1,1) S \rangle &= \frac{1}{\sqrt{2}} \left( \cre{c}_{{\rm L} \uparrow}
\cre{c}_{{\rm R} \downarrow} - \cre{c}_{{\rm L} \downarrow} \cre{c}_{{\rm
    R} \uparrow} \right) | 0 \rangle,
\\
\label{eq:tplusdef}
|T_+ \rangle &= \cre{c}_{{\rm L} \uparrow} \cre{c}_{{\rm R} \uparrow} | 0 \rangle,
\\
\label{eq:tzerodef}
|T_0 \rangle &= \frac{1}{\sqrt{2}} \left( \cre{c}_{{\rm L}
  \uparrow}\cre{c}_{{\rm R} \downarrow} + \cre{c}_{{\rm L} \downarrow}
\cre{c}_{{\rm R} \uparrow} \right) | 0 \rangle,
\\
\label{eq:tminusdef}
|T_- \rangle &= \cre{c}_{{\rm L} \downarrow} \cre{c}_{{\rm R} \downarrow}
| 0 \rangle.
\end{align}

The orbital parts of the basis states with single occupancy in each
well, i.e.~the spin-singlet $|(1,1)S\rangle $ and the spin-triplet $|
T_{0,\pm} \rangle$, are given by
\begin{equation}
\label{eq:orbitals2electronssingly}
\Psi\subup{s}_{\pm}(\vctr{r}_1,\vctr{r}_2) = \frac{1}{\sqrt{2}} \left[
  \Phi\sub{L} (\vctr{r}_1) \Phi\sub{R}(\vctr{r}_2) \pm \Phi\sub{R}
  (\vctr{r}_1) \Phi\sub{L}(\vctr{r}_2) \right],
\end{equation}
while the orbital parts of the two states $|(0,2)S \rangle$ and
$|(2,0)S \rangle$ with double occupation of the right and left wells,
respectively, are given by
\begin{equation}
\label{eq:orbitals2electronsdoubly}
\Psi\subup{d}\sub{L,R} (\vctr{r}_1,\vctr{r}_2) = \Phi\sub{L,R} (
\vctr{r}_1 )\Phi\sub{L,R} ( \vctr{r}_2 ).
\end{equation}
The orbital functions $\Psi\subup{s}_+$ and $\Psi\subup{d}\sub{L,R}$ are
symmetric under exchange of particles, and therefore must be
associated with  the antisymmetric singlet spin wave function (total
spin $S=0$), while $\Psi\subup{s}_-$ is antisymmetric under exchange,
and is associated with the symmetric triplet spin wave function (total
spin $S=1$).

The electrostatic gates that create the potential $V\sub{DQD}$ can, in
addition, tune the energies of the electrons in the potential minima
by creating an additional bias potential $V\sub{bias}$.  We model this bias
as a simple detuning $\varepsilon$, which gives an energy difference for an
electron occupying the left or the right dot, 
\begin{equation}
\label{eq:vbfielddef}
\varepsilon = \langle \Phi\sub{L} | eV\sub{bias} |
\Phi\sub{L} \rangle - \langle \Phi\sub{R} |
eV\sub{bias} | \Phi\sub{R} \rangle .
\end{equation}
In the symmetric case, $\varepsilon=0$, the voltages on the
electrostatic gates are set so that, in the absence of
electron-electron interactions, an electron would have the same
energy in either well.  The Coulomb repulsion,
Eq.(\ref{eq:coulombinteraction}),  penalizes the states $|(0,2)S
\rangle$ and $|(2,0)S \rangle$ with double occupation of either well
by an amount $U$, given by
\begin{equation}
\label{eq:udefbld}
U = \langle \Psi\subup{d}\sub{L,R} | C | \Psi\subup{d}\sub{L,R}
\rangle.
\end{equation}
Therefore, for a symmetric potential, $\varepsilon \approx 0$, the
lowest energy states of two electrons will be primarily comprised of
singly occupied orbitals.  When the detuning is large enough to
overcome the on-site electron-electron repulsion in one well,
$|\varepsilon| > U$, the doubly occupied state with both electrons on
the dot with lower potential becomes the ground state.   Varying the
gate voltages to increase the detuning from large and negative to
large and positive values then tunes the occupation numbers of the two
dots in the ground state of the DQD through the sequence
$(2,0)\rightarrow (1,1) \rightarrow (0,2)$.  Since the states with the
charge configurations $(2,0)$ and $(0,2)$ are  singlets, while those
with the $(1,1)$ charge configuration can be either singlet or
triplet,  the  measurement of charge as a function of detuning can
reveal the spin states.  

The strong  spin-independent interactions of the electrons with the
confinement potential, $V\sub{DQD}$, Coulomb repulsion $C$, as well
as the kinetic energy are, at low energies and in the limit of tight
confinement, described by the matrix elements between
Slater-determinant-type states in which the two electrons are loaded
into some combination of the the Wannier orbitals (see
Eq.(\ref{eq:udefbld}) and Ref.~(\cite{BLD99})):
\begin{align}
\label{eq:thdefbld}
t &=  \langle \Phi\sub{L,R} | h_{1,2}^0 |
\Phi\sub{R,L}\rangle - \frac{1}{\sqrt{2}} \langle \Psi\subup{s}_{+} |
C | \Psi\subup{d}\sub{L,R} \rangle,
\\
\label{eq:vpmdefbld}
V_{\pm} &= \langle \Psi\subup{s}_{\pm} | C |
\Psi\subup{s}_{\pm} \rangle,
\\
\label{eq:xdefbld}
X &= \langle \Psi\subup{d}\sub{L,R} | C | \Psi\subup{d}\sub{R,L}
\rangle.
\end{align}
Here, we have used $h^0_{1,2}= H\sub{osc} + V\sub{DQD}(\vctr{r}) -
V_{\rm h}(\vctr{r} \mp a \vctr{e}_{\xi})$ to label the part of
Hamiltonian that includes the kinetic energy $T$ and the harmonic part
of the potential $V\sub{DQD}$ near the dot centers.  The dots are
displaced by $\pm a$ along the axis with the unit vector
$\vctr{e}_{\xi}$, see Fig.~\ref{fig:axes}.  Thus,
$h^0_{1,2}-H\sub{osc}$ describes the tunneling due to the mismatch
between the true double-dot potential
$V\sub{DQD}$ and the potential of the harmonic wells \cite{BLD99}.
The matrix element $X$ describes coordinated hopping of two electrons
from one quantum dot to the other, $t$ is the renormalized single
electron hopping amplitude between the two dots, which includes
contributions of both the single-particle tunneling amplitude and the
Coulomb interaction, and $V_+$ ($V_-$) is the Coulomb energy in the
singlet (triplet) state with one electron in each well.

The confinement potential $V\sub{DQD}$, the Coulomb interaction $C$,
and the detuning $\varepsilon$ provide the largest energy scales in a
DQD.   These terms add  up to the spin-independent Hamiltonian $H_0 =
T + V\sub{DQD} + C + V\sub{bias}$, which within the space of the
six lowest energy dot orbitals, using the basis defined in
Eq.~(\ref{eq:20singletdef}) - Eq.~(\ref{eq:tminusdef}), is represented
by
\begin{equation}
\label{eq:hspinindependent}
H_0 = \left(
\begin{array}{cc}
H\sub{SS}
&
0
\\
0
&
H\sub{TT,0}
\end{array}
\right),
\end{equation}
where the singlet Hamiltonian in the basis $|(0,2)S \rangle$, $|(2,0)S
\rangle$, $|(1,1)S\rangle$ is
\begin{equation}
\label{eq:hess}
H\sub{SS} =
\left(
\begin{array}{ccc}
U-\varepsilon
&
X
&
- \sqrt{2} t
\\
X
&
U + \varepsilon
&
- \sqrt{2} t
\\
- \sqrt{2} t
&
- \sqrt{2} t
&
V_+
\end{array}
\right),
\end{equation}
and the triplet Hamiltonian is diagonal, $H\sub{TT, 0} = V_-$.

In addition to the terms described above, there are three sources of
spin-dependent interactions: Zeeman coupling to an external magnetic
field, hyperfine coupling between electron spins and nuclear spins in
a quantum dot, and the spin orbit interaction.  For now we neglect the
spin orbit interaction, and analyze it in detail in  the next section.

The direct coupling of the electron spins to a uniform external
magnetic field $\vctr{B}$ is described by the Zeeman term 
\begin{equation}
\label{eq:zeeman1}
H\sub{Z} = -g\mu_e \vctr{B} \cdot \left( \vctr{S}\sub{L} +
\vctr{S}\sub{R} \right),
\end{equation}
where $g$ is the electron $g-$factor and $\mu_e$ is the electron
magnetic moment.  In addition, the Fermi contact hyperfine interaction
between electron and nuclear spins reads
\begin{equation}
\label{eq:hnuclear}
H\sub{nuc} = \sum_i \vctr{h}_i \cdot \vctr{S}_i,
\end{equation}
where $\vctr{h} _i$, $i=L,R$, is the Overhauser field of the quantum
dot $i$, given by \cite{O53} 
\begin{equation}
\label{eq:overhauserDef}
\vctr{h}_i =  \sum_{j} A_j |\Psi_{i}(\vctr{R}_j)|^2 \vctr{I}_j.
\end{equation}
Here $A_j$ is the hyperfine coupling constant for the nuclear species
at site $j$, with typical size of the order of $90\mu\rm{eV}$ for GaAs
\cite{SKL03}, $\Psi_i$ is  the electron orbital envelope wave
function in the right ($i=R$) and left ($i=L$) dot, $\vctr{R}_j$ is
the position of the $j$th nucleus in the quantum dot, and $\vctr{I}_j$
is the corresponding nuclear spin.

\begin{figure}[t]
\begin{center}
\includegraphics[width=8.5cm]{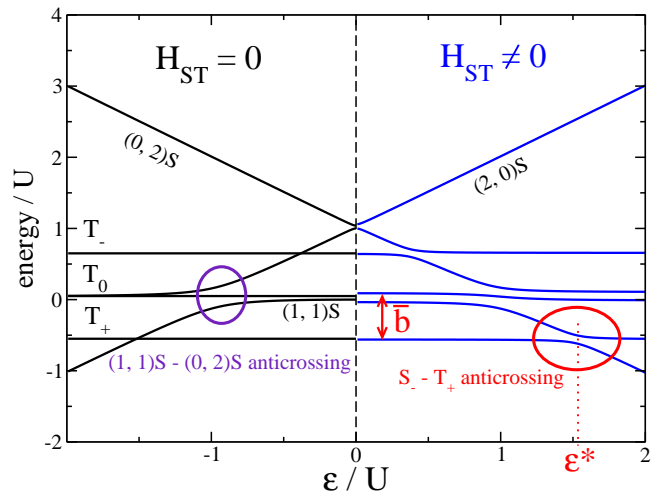}
\caption{\label{fig:cross02}Energy levels of the double quantum dot
  system obtained from exact numerical diagonalization of $H$ given in
  Eq.~(\ref{eq:hallmatrixoriginal}) and plotted as a function of the
  detuning $\varepsilon$ measured in units of the Coulomb on-site
  repulsion energy $U$. Of particular interest here are the crossings
  and anti-crossings of singlet and triplet states due to spin orbit
  and hyperfine interactions.  For $H\sub{ST}=0$ (see
  Eq.~(\ref{eq:hfullblock})), i.e. vanishing singlet-triplet mixing
  (left-hand side of plot), the parameter values chosen are $(U, t, p,
  X, V_-, V_+, \bar{b}, \delta b_{y}, \delta \vctr{b} \cdot \vctr{e},
  \delta \vctr{b} \cdot \vctr{e'}, \varphi ) = ( 1, 0.1, 0, 0, 0.05,
  0.04, 0.3, 0, 0, 0, 0)$.  In this case, the singlets
  $|(1,1)S\rangle$ and $|(0,2)S\rangle$ anti-cross (left oval) and a
  finite gap  opens, whereas the singlets and triplets only cross (no
  gap).   For $H\sub{ST}\neq 0$  additional gaps open (right-hand side
  of plot), in particular at the lower singlet-triplet anti-crossing
  around $\varepsilon=\varepsilon^*$ (right oval) with an energy
  splitting  $\Delta_{ST}$ that depends on magnetic field, detuning,
  spin orbit  and hyperfine interactions (see main text and figures
  below).  The singlet $S_{-}$ is a superposition of $|(1,1)S\rangle$
  and $|(0,2)S\rangle$ (see Eq.~(\ref{eq:spmdef})).  The parameter
  values chosen for the right plot are $(U, t, p, X, V_-, V_+,
  \bar{b}, \delta b_{y}, {\delta{\vctr{b}} \cdot \vctr{e}}, \delta
  \vctr{b} \cdot {\vctr{e'}}, \varphi) = (1, 0.1, 0.01,  0, 0.05,
  0.04, 0.3, 0.02, 0.02, 0.01, \pi/2)$. }
\end{center}
\end{figure}

Because the Zeeman and hyperfine interactions, Eqs.~(\ref{eq:zeeman1})
and (\ref{eq:hnuclear}), have similar forms, we combine them into a
single effective field that acts on the electron spin in each dot:
\begin{equation}
\label{eq:hbeff0}
H\sub{nuc} + H\sub{Z} = -\vctr{b}\sub{L} \cdot \vctr{S}\sub{L} -
\vctr{b}\sub{R} \cdot \vctr{S}\sub{R}.
\end{equation}
The effective fields $\vctr{b}\sub{L,R} = g \mu\sub{B}
\vctr{B}\sub{L,R} - \vctr{h}\sub{L,R}$ include the contributions of
the external and Overhauser fields, with all coupling constants
absorbed in the field definitions.     The energy levels arising from
the spin-conserving Hamiltonian, Eq.~(\ref{eq:hspinindependent}), along
with coupling to a uniform effective field, Eq.~(\ref{eq:hbeff0}) with
$\vctr{b}\sub{L} = \vctr{b}_R$, are shown in the left panel of
Fig.~\ref{fig:cross02}.

Below we will be interested in transitions that change the total spin of
the pair of electrons in the DQD.  To facilitate the discussion, we
separate the total field into a sum-field $\bar{\vctr{b}} =
(\vctr{b}\sub{L} + \vctr{b}\sub{R})/2$ and a difference field
$\delta\vctr{b} = (\vctr{b}\sub{L} - \vctr{b}\sub{R})/2$:
\begin{equation}
\label{eq:hbeff}
H\sub{nuc} + H\sub{Z}  = -\vctr{\bar{b}} \cdot \left( \vctr{S}\sub{L}
+ \vctr{S}\sub{R} \right) - \delta\vctr{b} \cdot \left(
\vctr{S}\sub{L} - \vctr{S}\sub{R} \right).
\end{equation}
The symmetric component $\bar{\vctr{b}}$ conserves the magnitude of
the total spin, $\left[ \bar{\vctr{b}}, \left( \vctr{S}\sub{L} + \vctr{S}\sub{R}
  \right)^2 \right] =0$, while the antisymmetric component
$\delta\vctr{b}$ does not.  We include the spin-conserving field
$\bar{\vctr{b}}$ into the unperturbed Hamiltonian, and define 
\begin{equation}
\label{eq:httdef}
H\sub{TT} = H_{TT,0} - \bar{\vctr{b}} \cdot \left( \vctr{S}\sub{L} +
\vctr{S}\sub{R} \right).  
\end{equation}

Below we will investigate the role of the Overhauser  fields in
causing spin transitions near a singlet-triplet level crossing in a
two-electron DQD.   While the external magnetic field $\vctr{B}$ is a
classical variable, the Overhauser fields $\vctr{h}\sub{L,R}$ are, in
principle, quantum operators that involve a large number of nuclear
spins, see Eq.~(\ref{eq:overhauserDef}).   Hyperfine-induced electron
spin transitions may be accompanied by nuclear spin flips, and the
dynamical, quantum nature of the Overhauser field may be important.
However, due to the large number of nuclear spins, the time scale for
the Overhauser fields $\vctr{h}\sub{L,R}$ to change appreciably can be
much longer than the time spent near the avoided crossing, where spin
transitions are possible.  Thus we will treat the fields
$\bar{\vctr{b}}$ and $\delta\vctr{b}$ as quasi-static classical
variables, including a discussion of the averaging which occurs due to
nuclear Larmor precession and statistical (thermal) fluctuations.

\section{Spin orbit interaction}
\label{sec:soi}

In addition to the external and hyperfine fields, electron spins in a
DQD are also influenced by orbital motion due to the spin-orbit
interaction.  Here we describe how the bulk spin-orbit coupling of the
2DEG is manifested in the confined DQD system.  In GaAs quantum wells,
the spin-orbit interaction is caused by the inversion asymmetry of the
interface that forms the quantum well \cite{R60,BR84,DK86} and the
inversion asymmetry of host material \cite{D55}.  With the 2DEG being
much thinner than the lateral quantum dot dimensions, both spin orbit
interactions are linear in the in-plane momenta of the confined
electrons, and together are given by
\begin{equation}
\label{eq:so1}
H\sub{SO} = \alpha \left(p_{x'}\sigma _{y'} - p_{y'}\sigma _{x'}
\right) + \beta \left( - p_{x'} \sigma _{x'} + p_{y'} \sigma _{y'}
\right),
\end{equation}
where the Rashba and  Dresselhaus spin orbit interaction constants
$\alpha$ and $\beta$, respectively, depend on the thickness and shape
of the confinement in the growth direction and on the material
properties of the heterostructure in which the 2DEG is fabricated.
This form of spin-orbit coupling appears in a quantum well fabricated
in the $(001)$ plane of GaAs crystal, and the $x'$- and $y'$-axes
point along the crystallographic directions $[100]$ and $[010]$,
respectively.

Within the space of low-energy single-electron orbitals in the DQD,
the action of the spin orbit interaction can be expressed in terms of
a spin-orbit field  $\boldsymbol{\Omega}$, 
\begin{equation}
\label{eq:soidef2nd}
H\sub{SO} = \frac{i}{2} \boldsymbol{\Omega} \cdot \sum_{\alpha,\beta =
  \uparrow \downarrow} \left( \cre{c}_{ {\rm{L}} \alpha}
\boldsymbol{\sigma}^{\alpha \beta} \ann{c}_{ {\rm{R}} \beta} - {\rm
  h.c.} \right),
\end{equation}
where the field
\begin{equation}
\label{eq:omegadef}
i\vctr{\Omega}= \langle \Phi\sub{L} | p_{\xi} | \Phi\sub{R} \rangle
{\boldsymbol{a}_{\vctr{\Omega}}}
\end{equation}
depends on the orientation of the dots with respect to
crystallographic axes through the vector
$\boldsymbol{a}_{\vctr{\Omega}}$ \cite{SBD+03}.  For a 2DEG in the
$(001)$ plane, $\boldsymbol{a}_{\vctr{\Omega}}$ is given by
\begin{equation}
\label{eq:etadefsbd+}
\vctr{a}_{\vctr{\Omega}} = \left( \beta - \alpha \right) \cos \theta
\vctr{e}_{[\bar{1}10]} + \left( \beta + \alpha \right) \sin \theta
\vctr{e}_{[110]},
\end{equation}
where the angle between the $\vctr{e}_{\xi}$ direction and the $[110]$
crystallographic axis is denoted by $\theta$.   The matrix element of
$p_{\xi}$, the momentum component along the $\xi$-direction that
connects the two dots, see Fig.~\ref{fig:axes}, is taken between the
corresponding Wannier orbitals, and it depends on the envelope wave
function and the double dot binding potential.  The spin orbit field
$\vctr{\Omega}$ accounts for the spin rotation when the electron hops
between the dots.  Therefore, the spin orbit interaction enables
transitions between triplet states with single occupation of each
well, to the singlet states with double occupation of either the left
or the right well. 

The matrix element in Eq.~(\ref{eq:omegadef}) can be calculated
explicitly in a model potential \cite{SBD+03,BLD99}, giving
\begin{equation}
\label{eq:pres}
\vctr{\Omega} = \frac{4t}{3} \frac{l}{\Lambda\sub{SO}}
\frac{\boldsymbol{a}_{\vctr{\Omega}}}{|\boldsymbol{a}_{\vctr{\Omega}}|},
\end{equation}
where $l$ is the interdot distance.  The numerical prefactor is
model-dependent, but the dependence on other parameters is generic.
The hopping amplitude $t$, and the interdot separation $l$ depend on
the geometry of the double dot system, whereas the spin-orbit length
$\Lambda\sub{SO}$ is determined by material properties (Rashba and
Dresselhaus spin orbit strength) and by the orientation of the DQD
with respect to the crystallographic axes.  In particular, if the 2DEG
lies in the $(001)$ plane, it is given by 
\begin{equation}
\label{eq:lambdalambdapm}
\frac{1}{\Lambda\sub{SO}} = \sqrt{ \left(
  \frac{\cos\theta}{\lambda_-}\right)^2 +  \left(
  \frac{\sin\theta}{\lambda_+} \right)^2 },
\end{equation}
where $\lambda_{\pm} = \hbar/[m^*(\beta \pm \alpha)]$ \cite{GKL08},
with $m^*$ being the effective band mass of the electron.  In the
special case $\beta=0$, $\theta=0$, this definition reduces to the
Rashba spin orbit length $\Lambda\sub{SO}\left|_{\beta=\theta=0}\right. =
\lambda\sub{SO}=\hbar/m^*\alpha$.

One of the main goals in the following is to derive the dependence of
the energy splitting at the anticrossing between the lowest-energy
electron spin triplet and singlet states (see Fig.~\ref{fig:cross02},
$S_- - T_+$ anticrossing) in terms of this spin-orbit length
$\Lambda\sub{SO}$.  Detailed understanding of this dependence may then
be used to extract the value of $\Lambda_{\rm SO}$, e.g. from
measurements of the singlet-triplet transition probability in
gate-sweep experiments. 

Within the model used above for explicit calculation, the components
of $\boldsymbol{\Omega}$ are real, even in the presence of magnetic
fields.   This is due to the high symmetry of the ground state
orbitals of the quantum dot in the model, and remains true even after
the replacement $p_{\xi} \rightarrow p_{\xi} - (e/c) A_{\xi}$ in the
spin-orbit Hamiltonian $H\sub{SO}$, Eq.~(\ref{eq:so1}).   However,
this fact is not essential for the physics described below.

\section{Singlet-triplet transitions and the choice of spin
  quantization axis} \label{sec:eh}

Transitions between singlet and triplet states can be mediated by the
spin orbit interaction or by an inhomogeneous effective magnetic field
(external plus hyperfine), $\delta \vctr{b}$.  In \cite{RNL+10} it was
shown that the transfer of angular momentum between electrons and the
nuclei strongly depends on the relative size and phase of the electron
spin flip matrix elements induced by spin orbit interaction and by the
difference of the Overhauser fields in the two dots.  Using our model
of a detuned DQD, we will study these matrix elements in the following
in detail and in particular focus on the singlet-triplet level
splitting, see Fig.~\ref{fig:cross02}, right panel.

The homogeneous field $\bar{\vctr{b}}$ acts only within the spin
triplet subspace, while the inhomogeneous field $\delta\vctr{b}$ mixes
singlet $S=0$ and triplet $S=1$ states.  Representing the total
Hamiltonian in the basis $\{(|(0,2)S\rangle, |(2,0)S\rangle,
|(1,1)S\rangle, |T_{z+}\rangle, |T_{z0}\rangle, |T_{z-}\rangle)\}$,
where the $z$-axis is taken along $\boldsymbol{\Omega}$, see
Fig.~\ref{fig:axes}, we find
\begin{widetext}
\begin{equation}
\label{eq:hallmatrixoriginal}
H= \left(
\begin{array}{cccccc}
U-\varepsilon
&
X
&
-\sqrt{2} t
&
0
&
-i \sqrt{2} \Omega
&
0
\\
X
&
U+\varepsilon
&
-\sqrt{2} t
&
0
&
-i\sqrt{2} \Omega
&
0
\\
-\sqrt{2} t
&
-\sqrt{2} t
&
V_+
&
-\sqrt{2} \left( \delta b_{x} - i \delta b_{y} \right)
&
2 \delta b_{z}
&
\sqrt{2} \left( \delta b_{x} + i \delta b_{y} \right)
\\
0
&
0
&
-\sqrt{2} \left( \delta b_{x} + i \delta b_{y} \right)
&
V_- + 2 \bar{b}_{z}
&
\bar{b}_{x} \sqrt{2}
&
0
\\
i \sqrt{2} \Omega
&
i \sqrt{2} \Omega
&
2 \delta b_{z}
&
\bar{b}_{x} \sqrt{2}
&
V_-
&
\bar{b}_{x} \sqrt{2}
\\
0
&
0
&
\sqrt{2} \left( \delta b_{x} - i \delta b_{y} \right)
&
0
&
\bar{b}_{x} \sqrt{2}
&
V_- - 2 \bar{b}_{z}
\end{array}
\right).
\end{equation}
\end{widetext}

The Hamiltonian $H$, Eq.~(\ref{eq:hallmatrixoriginal}), is the
starting point for all of our further calculations.  Our results will
show the dependence of the singlet-triplet splitting on the parameters
that enter $H$.  This Hamiltonian describes a double quantum dot with
single orbital per dot, i.e.~in the Hund-Mulliken approximation, and it is
valid as long as the dot quantization energy is the largest energy
scale in the problem.  It can be applied to double quantum dots of
various kinds, for example the gated lateral or vertical dots in III-V
semiconductor materials, quantum dots in nanowires, or self assembled
quantum dots.  We illustrate the spectrum of $H$ in
Fig.~\ref{fig:cross02}, for a set of parameters that emphasizes
anticrossings of the levels.  The spectrum is obtained by exact
diagonalization of $H$, and it is given as a function of detuning
$\epsilon$.  For other types of quantum dots, the parameter values
would change, but the overall structure of the spectrum remains the
same.

\section{Effective Hamiltonian near the singlet-triplet anticrossing}
\label{sec:heff}

In the limit of large detuning, $|\varepsilon| \gg U, V_+, V_-,
|\bar{\vctr{b}}|$, the ground state is a spin singlet with both
electrons in either the left or right dot, depending on whether
$\varepsilon > 0$ or $\varepsilon < 0$.  In the region of weak
detuning, the ground state has one electron in each of the dots.  For
a sufficiently strong sum-field, $|\bar{\vctr{b}}| > U - V_+$, the
singlet ground state exhibits an avoided level crossing with the
lowest energy triplet state, i.e. the $S=1$ state oriented along
$\bar{\vctr{b}}$, see Fig.~\ref{fig:axes},  at a detuning where the
potential energy gained by the singlet's double occupancy of the lower
well compensates the Zeeman energy gained by the spin-polarized
triplet.   Here, the residual splitting  is determined by the spin
non-conserving interactions.  The behavior near this anticrossing has
been the focus of many recent studies on the interaction of electron
spins with the nuclei \cite{RTP+08,RTL+08,BRM+09,BFM+10,SPJ+11}.
The role of spin orbit interaction has received less attention than
that of the nuclei, and will be analyzed in the following sections.

The orbital structure of the levels near the anticrossing is
determined by the spin-independent interactions and by the direction
and amplitude of the sum-field $\bar{\vctr{b}}$.  The singlet subspace
acted on by the Hamiltonian in Eq.~(\ref{eq:hess}) includes a state
with single occupation of the two dots, and two states which feature
double occupation of either the left or the right dot.  Generically,
the state that takes part in the anticrossing includes amplitudes of
all three singlet states.  However, because $U \gg J$, where $J\approx
4t^2/(U - V_+) \sim 0.01 - 0.1 \unit{meV}$ is the splitting between
the lowest-energy triplet and the lowest-energy singlet state, and
$U\sim 1\unit{meV}$, admixture of at least one of the
singlets will always be suppressed at the anticrossing by a large
energy denominator  (note that $ t\approx 0.01 - 0.1 \unit{meV} \ll U
-V_+ )$. 

Let us now construct the effective Hamiltonian which acts in the
two-level subspace spanned by the levels near the anticrossing.
First, the spin-conserving part of the full (6$\times$6) Hamiltonian
reads
\begin{equation}
\label{eq:h1120spinconservingblocks}
H\sub{sc} =
\left(
\begin{array}{cc}
H\sub{SS} & 0
\\
0 & H\sub{TT}
\end{array}
\right),
\end{equation}
where the block $H\sub{SS}$ acts in the singlet subspace, $H\sub{TT}$
acts in the triplet subspace, and the off-diagonal blocks vanish due
to spin conservation.

Explicitly, the block $H\sub{SS}$ is given by Eq.~(\ref{eq:hess}) in
the  basis Eq.~(\ref{eq:20singletdef})-(\ref{eq:tminusdef}).  The
triplet block $H\sub{TT}$ is given in Eq.~(\ref{eq:httdef}).  Since
$U\gg t \gg X$ in a typical quantum dot, the state at the anticrossing
can at most include significant contributions from two out of three
basis singlets.  We will consider the anticrossing at positive values
of the detuning $\varepsilon$ (the anticrossing at negative voltage is
analogous) so that the $|(2,0)S\rangle$ state with the energy
$U+\varepsilon$ is far detuned from the other two singlets.  The
remaining two singlets can be close in energy.  In order to include
the possibility of near degeneracy, we will introduce a mixing angle
$\psi$ that parametrizes the hybridization of the  $|(0,2)S\rangle$
and $|(1,1)S\rangle$ states.  In the restricted subspace of these
hybridized states, the singlet Hamiltonian reads: 
\begin{equation}
\label{eq:hsssplitting}
H\sub{r} = \frac{ U - \varepsilon + V_+ }{2} + \boldsymbol{\tau} \cdot
\vctr{n} \sqrt{ \frac{( U - \varepsilon - V_+ )^2}{4} + 2 t^2 },
\end{equation}
where $\boldsymbol{\tau} = (\tau_x, \tau_y, \tau_z)$ is the vector of
Pauli matrices. The  pseudospin $\boldsymbol{\tau}$ describes the
components of the anticrossing singlet state, $|\tau_z = 1\rangle =
|(1,1)S \rangle$, $|\tau_z = -1\rangle = |(0,2)S\rangle$, within the
approximation that we neglect the remaining $|(2,0)S\rangle$ component
(this is valid when $|t/U|\ll 1$).  In this case, $\vctr{n} =
\vctr{e}_z \cos{2\psi}  + \vctr{e}_x \sin{2\psi}$ is a unit vector
parametrized by the mixing angle $\psi$ that describes the relative
size of mixing and splitting of $\tau_z$ eigenstates.  Note that the
$y$-component of $\vctr{n}$ vanishes due to the choice of phases in
the quantum dot ground states, which guarantees that the
spin-independent hopping matrix element $t$ is purely real.   We
remark again that in our DQD set-up the hopping matrix element $t$
stays real even in the presence of magnetic fields, \cite{BLD99}.  The
mixing angle of the doubly and singly occupied  states at the $S=0$
anticrossing, $|(0,2)S\rangle$ and $|(1,1)S\rangle$, respectively, is
defined by 
\begin{align}
\label{eq:psidefcos}
\cos{2\psi} &= \frac{U - V_+ - \varepsilon}{\sqrt{ \left( U - V_+
    -\varepsilon \right) ^2 + 8 t^2 }}, 
\\
\label{eq:psidefsin}
\sin{2\psi} &= \frac{ 2 \sqrt{2} t}{\sqrt{ \left( U - V_+ -
    \varepsilon \right)^2 + 8 t^2 }}. 
\end{align}

In the basis of eigenstates of Eq.~(\ref{eq:hsssplitting}) the singlet
Hamiltonian $H\sub{SS}$ is given by
\begin{widetext}
\begin{equation}
\label{eq:hss1120spinconserving}
H\sub{SS} =
\left(
\begin{array}{ccc}
U + \varepsilon
&
X \cos\psi - \sqrt{2} t \sin\psi
&
-X \sin\psi - \sqrt{2} t \cos\psi
\\
X \cos\psi - \sqrt{2} t \sin\psi
&
E_{S+}
&
0
\\
-X \sin\psi - \sqrt{2} t \cos\psi
&
0
&
E_{S-}
\end{array}
\right),
\end{equation}
\end{widetext}
where 
\begin{equation}
\label{eq:espmdef}
E_{S\pm} = \frac{ U-\varepsilon + V_+}{2} \pm
\sqrt{(U-\varepsilon-V_+)^{2}/4 + 2t^2} 
\end{equation}
are the eigenvalues  of $H\sub{SS}$ in the spin-conserving sector.
The  basis vectors used here are the far-detuned singlet
$|(2,0)S\rangle$  (Eq.~(\ref{eq:20singletdef})) and the singlets
$|S_{\pm}\rangle$ defined by
\begin{align}
\label{eq:spmdef}
|S_+\rangle &=  \sin \psi |(1,1)S\rangle - \cos\psi |(0,2)S\rangle,
\\
|S_-\rangle &= \cos \psi |(1,1)S\rangle + \sin \psi |(0,2)S\rangle.
\end{align}
In the limit $U\gg t$, $|S_\pm\rangle$ become eigenstates of
$H\sub{SS}$ with energies $E\sub{S\pm}$ given in
Eq.~(\ref{eq:espmdef}).

While the mixing of  orbital states belonging to singlets does not
affect the triplet Hamiltonian $H\sub{TT}$, it will change the form of
coupling between the singlet and triplet states near the anticrossing.
In the following, we first diagonalize the triplet sector in order to
find the explicit form of the  triplet state at the anticrossing, and
then find the effective Hamiltonian of the singlet-triplet coupling.

The triplets are Zeeman split by the sum-field $\bar{\vctr{b}}$.  We
have chosen the $z$-axis of spin quantization so that the spin-orbit
interaction couples $|(0,2)S \rangle$ and $|(2,0)S\rangle$ to the
$|S=1, S_z=1 \rangle$ state.  We will now diagonalize the triplet part
of the spin-conserving Hamiltonian, given by 
\begin{equation}
\label{eq:htriplet}
H\sub{TT} = V_- + 2 \bar{b}
\left(
\begin{array}{ccc}
-\cos \varphi
&
\frac{1}{\sqrt{2}}\sin \varphi
&
0
\\
\frac{1}{\sqrt{2}} \sin \varphi
&
0
&
\frac{1}{\sqrt{2}} \sin \varphi
\\
0
&
\frac{1}{\sqrt{2}} \sin \varphi
&
\cos \varphi
\end{array}
\right),
\end{equation}
where we have used $\bar{b} = | \bar{\vctr{b}} |$, $\cos \varphi =
\bar{b}_{z} / \bar{b}$, and $\sin \varphi = \bar{b}_{x} /\bar{b}$ (see
Fig.~\ref{fig:axes}).   The unitary transformation $\ann{U}\sub{t}
H\sub{TT} U^{\dagger}\sub{t}$ that diagonalizes $H\sub{TT}$ is
\begin{equation}
\label{eq:utriplet}
U\sub{t} = \left(
\begin{array}{ccc}
\cos ^2 \frac{\varphi}{2}
&
-\frac{1}{\sqrt{2}} \sin \varphi
&
\sin ^2 \frac{\varphi}{2}
\\
-\frac{1}{\sqrt{2}} \sin \varphi
&
-\cos \varphi
&
\frac{1}{\sqrt{2}} \sin \varphi
\\
\sin ^2 \frac{\varphi}{2}
&
\frac{1}{\sqrt{2}} \sin \varphi
&
\cos ^2 \frac{\varphi}{2}
\end{array}
\right).
\end{equation}
We denote the basis states by  $|T_+\rangle$, $|T_0\rangle$, and
$|T_-\rangle$, where, now, the quantization axis is given by the
sum-field $\bar{\vctr{b}}$.

With the diagonalization of the triplet block of the spin-conserving
Hamiltonian, and the preceding approximate diagonalization of the
spin-conserving singlet Hamiltonian,
Eq.~(\ref{eq:hss1120spinconserving}), we are able to describe the
spin-conserving interaction near the anticrossing in a convenient
form.  We will use
$|S_-\rangle$ and $|T_+\rangle$ as basis vectors, and study the
effective Hamiltonian in the vicinity of the anticrossing that can
emerge from spin non-conserving interactions.

The full Hamiltonian is 
\begin{equation}
\label{eq:hschsnc}
H=H\sub{sc} + H\sub{SO} - \delta{\vctr{b}} \cdot \left(
\vctr{S}\sub{L} - \vctr{S}\sub{R} \right), 
\end{equation}
and reads in block form
\begin{equation}
\label{eq:hfullblock}
H=\left(
\begin{array}{cc}
H\sub{SS}
&
H\sub{ST}
\\
H\sub{TS}
&
H\sub{TT}
\end{array}
\right).
\end{equation}
The diagonal blocks $H\sub{SS}$ and $H\sub{TT}$ give the
spin-conserving part, denoted by $H\sub{sc}$, while the off-diagonal
blocks $H\sub{ST}$ and $H\sub{TS}=H\sub{ST}^{\dagger}$ induce
singlet-triplet transitions.  In the basis $(|(2,0)S\rangle,
|S_+\rangle, |S_-\rangle, |T_+\rangle, |T_0\rangle, |T_-\rangle)$,
the diagonal blocks take a simple form.  The singlet block is
$H\sub{SS}$ given in Eq.~(\ref{eq:hss1120spinconserving}).  The
triplet block $H\sub{TT}$ is diagonal and reads
\begin{equation}
\label{eq:tdiag}
H\sub{TT}={\rm diag} \left( V_- - |\bar{\vctr{b}}|, V_-, V_- +
|\bar{\vctr{b}}| \right), 
\end{equation}
where we used that $|T_+\rangle$  is the lowest-energy triplet.

The effective Hamiltonian near the anticrossing is determined by the
spin-conserving terms $t,U,X,\varepsilon,\bar{\vctr{b}}$, and by the
spin non-conserving terms $\boldsymbol{\Omega}$, arising from spin-orbit
coupling, and  $\delta \vctr{b}$, the effective difference field.
All  these interactions can be treated perturbatively in dots with
weak overlap of the orbitals.   The off-diagonal terms are denoted by
$H\sub{ST} = H\sub{ST}\subup{SO} + H\sub{ST}\subup{\delta b}$, and
$H\sub{TS}=H\sub{ST}^{\dagger}$, where
\begin{widetext}
\begin{equation}
\label{eq:hstso}
H\sub{ST}\subup{SO} = i \Omega
\left(
\begin{array}{ccc}
- \sin{\varphi}
&
- \sqrt{2}  \cos{\varphi}
&
\sin{\varphi}
\\
 \cos{\psi} \sin{\varphi}
&
\sqrt{2}  \cos{\psi} \cos{\varphi} 
&
- \cos{\psi} \sin{\varphi}
\\
- \sin{\psi} \sin{\varphi} 
&
- \sqrt{2}  \sin{\psi} \cos{\varphi} 
&
 \sin{\psi} \sin{\varphi} 
\end{array}
\right),
\end{equation}
\begin{equation}
\label{eq:hstdb}
H\sub{ST}\subup{\delta {\vctr{b}}} = 
\left(
\begin{array}{ccc}
0
&
0
&
0
\\
 \sqrt{2} \left( i\delta b_{y} + \delta \vctr{b} \cdot \vctr{e'}
 \right) \sin \psi 
&
2  \left( \delta \vctr{b} \cdot \vctr{e} \right) \sin \psi
&
\sqrt{2}  \left( i \delta b_{y} - \delta\vctr{b} \cdot \vctr{e'}
\right) \sin \psi 
\\
\sqrt{2} \left( i \delta b_y + \delta \vctr{b} \cdot \vctr{e'} \right)
\cos \psi 
&
2  \left( \delta \vctr{b} \cdot \vctr{e} \right) \cos \psi
&
\sqrt{2}  \left( i \delta b_y - \delta \vctr{b} \cdot \vctr{e'}
\right) \cos \psi  
\end{array}
\right),
\end{equation}
\end{widetext}
where the unit vector
\begin{equation}
\label{eq:edef}
\vctr{e} = \vctr{e}_{x} \sin \varphi + \vctr{e}_z \cos \varphi
\end{equation}
points in the direction of the homogeneous field $\bar{\vctr{b}}$, and
the vector
\begin{equation}
\label{eq:eprimedef}
\vctr{e'} = - \vctr{e}_x \cos\varphi + \vctr{e}_z \sin \varphi
\end{equation}
lies in the $xz$-plane, which contains ${\boldsymbol{\Omega}}$ and
$\bar{\vctr{b}}$, and points in the direction normal to $\bar{\vctr{b}}$.

In the vicinity of the anticrossing, the DQD behaves as an effective
two-level system, with the dynamics described by an effective
Hamiltonian denoted by $H\sub{cr}$.  Up to first order in spin
non-conserving interactions and, after neglecting the high-energy
state $|(2,0)S\rangle$, we find
\begin{equation}
\label{eq:hcr1}
H\sub{cr}^{(1)} =
\left(
\begin{array}{cc}
E_{S-} &  H_{34}
\\
H_{43} & E_{T+}
\end{array}
\right),
\end{equation}
where $H_{34} = \langle S_- | H | T_+\rangle$ is the matrix element of
$H$, Eq.~(\ref{eq:hfullblock}), between the anticrossing states, and
$E_{T+}=V_- - \bar{b}$.

\section{Singlet-triplet splitting at the anticrossing}
\label{sec:splitting}

The singlet-triplet splitting  $\Delta\sub{ST}$ at the $S_{-}-T_{+}$
anticrossing, see Fig.~\ref{fig:cross02}, can be accessed in the spin
blockade regime by transport measurements  or by charge sensing.  This
splitting gives valuable information about the properties of the
quantum dots and the nuclear polarization.  We shall derive now
explicit expressions for  $\Delta\sub{ST}$ in terms  of the
experimentally relevant quantities $\varepsilon$, $\bar{\vctr{b}}$,
$\delta \vctr{b}$, and $\vctr{\Omega}$.  We proceed by perturbation
expansion in $H\sub{ST}$, $t/U$, and $X/U$. First, we focus on the
first-order contributions and afterward address the higher order
corrections which become relevant around the points  where the leading
contributions can be tuned to zero by the control parameters.  

As the DQD detuning $\varepsilon$ is varied with other parameters held
fixed, there is a special value $\varepsilon^*$ where the energy
$E_{S-}$ of the lowest singlet, Eq.~(\ref{eq:espmdef}), becomes equal
to the energy $E_{T+}=V_- -\bar{b}$ of the triplet $|T+\rangle$,
Eq.~(\ref{eq:tdiag}).   The detuning at which this crossing occurs is
controlled by the amplitude $\bar{b}$ of the sum-field, as well as the
tunnel coupling $t$.   For the unperturbed case, described by $H_0$,
$\varepsilon^*$ is the solution of the equation
\begin{equation}
\label{eq:crossingpoint}
E\sub{S-}(\varepsilon^*) = V_- - \bar{b}.
\end{equation}
When the spin non-conserving interactions are taken into account, the
crossing of singlet ($S=0$) and triplet ($S=1$) states is avoided due
to state mixing (hybridization).

Up to first order in $H\sub{ST}$, $t/U$, and $X/U$, the splitting
follows from Eq.~(\ref{eq:hcr1}) and reads
\begin{equation}
\label{eq:deltadef}
\Delta\sub{ST}(\varepsilon,\bar{b},\varphi) = 2\sqrt{ |H_{34}|^2 +
  (E_{S-} - E_{T_+} )^2}.
\end{equation}
For a fixed value of $\bar{b}$, the splitting attains its minimum
value $\Delta^*\sub{ST}$ at $\varepsilon^*$, 
\begin{equation}
\label{eq:deltastar}
\Delta^*\sub{ST} \equiv \Delta\sub{ST}(\varepsilon^*(\bar{b}),
\bar{b},\varphi) = \min_{\varepsilon} \Delta\sub{ST}(\varepsilon,
\bar{b},\varphi),
\end{equation}
where $\varepsilon^*$ implicitly depends on $\bar{b}$, as well as
other DQD parameters.  From Eqs.~(\ref{eq:hcr1}), (\ref{eq:hstso}),
and (\ref{eq:hstdb}), the splitting is equal to $2|H_{34}|$,
\begin{equation}
\label{eq:deltast}
\Delta^*\sub{ST} = 2 \left| - i \Omega \sin \varphi \sin \psi  +
\sqrt{2} \left( \delta \vctr{b} \cdot \vctr{e'} + i \delta b_y \right) \cos\psi
\right|.
\end{equation}
Note that $\Delta^*\sub{ST}$ contains contributions from both the
spin-orbit coupling and the difference field $\delta \vctr{b}$.  The
relative importance of each of the two terms depends on the detuning
$\varepsilon^*$ through the mixing angle $\psi$, as well as on the
geometry through the angle $\varphi$ between the effective field
$\bar{\vctr{b}}$ and  the spin-orbit field $\vctr{\Omega}$.  When varying
the detuning $\varepsilon$ from large to small values, $\psi$
decreases from $\psi\approx \pi/2$ at strong detuning, $\varepsilon
\gg U -V_+$, to $\psi \approx 0$ at $|\varepsilon| \ll t$.  For a
mixing angle $\psi \approx \pi/2$,   the contribution to
$\Delta^*\sub{ST}$ coming from the spin orbit interaction  dominates
the one from the difference-field, and vice versa  for $\psi \approx
0$.  

Reaching the $\psi \approx 0$ regime requires weak magnetic fields,
i.e. $\bar{{b}}\ll t$, and in this case the energy splittings between
the triplet states are not large enough to make use of the simple model
for a two-level anticrossing.   On the other hand, reaching the regime
with $\psi \approx \pi/2$  requires that the detuning $\varepsilon^*$
at which $| S_- \rangle$ and $|T_+ \rangle$ anticross is far away from
$\varepsilon_{12}$, the detuning at the anticrossing of
$|(1,1)S\rangle$ and $|(0,2)S\rangle$ singlets,
cf. Fig.~\ref{fig:cross02}.  The width of the $|(1,1)S\rangle-|(0,2)S\rangle$
anticrossing is of the order $t$, so the requirement is
$|\varepsilon^* - \varepsilon_{12}| \gg t$.  Therefore, the Zeeman
energy of $|T_+\rangle$ must be larger than $t$, so $g\mu\sub{B} B \gg
t$, which gives $B \gg 0.2 \unit{T}$ for typical values $t \sim 10\unit{\mu
  eV}$ and $|g|=0.4$.  

These considerations show that, at least in principle, the relative
strengths of the spin-orbit and hyperfine contributions to the
singlet-triplet coupling can be tuned through a wide range of values
using a combination of gate voltages and magnetic field strength and
direction.  What situation do we expect for typical GaAs dots?  Using
a value of 3-5 mT for the random hyperfine field (see
e.g.~Ref.~\cite{HKP+07} and references therein), and an electron
$g$-factor $|g|=0.4$, we estimate $|\delta b|\approx 70-120$ neV.  For
the spin-orbit coupling $\Omega$, see Eq.(\ref{eq:pres}), using $t =
10 \mu eV$, an interdot separation $l = 50$ nm, and a spin-orbit
length $\Lambda_{\rm SO}$ in the range 6-30 $\mu$m (see
e.g.~Refs.\cite{ZMM+02, NKN+07}), we find $|\Omega|\approx 20-110$
neV.  Parameters may vary from device to device, but it appears that
the spin-orbit and hyperfine couplings are generally of similar orders
of magnitude, with the spin-orbit coupling typically a few times
weaker.  Thus adjustments of the matrix elements over a reasonable
range of $\psi$ may be sufficient to explore both the hyperfine and
spin-orbit dominated regimes.  Similar analysis can be performed for
devices in other materials, such as InAs or InSb nanowires, where the
natural balance between hyperfine and spin-orbit couplings may shift.

\subsection{Singlet-triplet splitting $\Delta\sub{ST}^*$ for $\vctr{\delta b} = 0$}

\begin{figure}[t]
\begin{center}
\includegraphics[width=8.5cm]{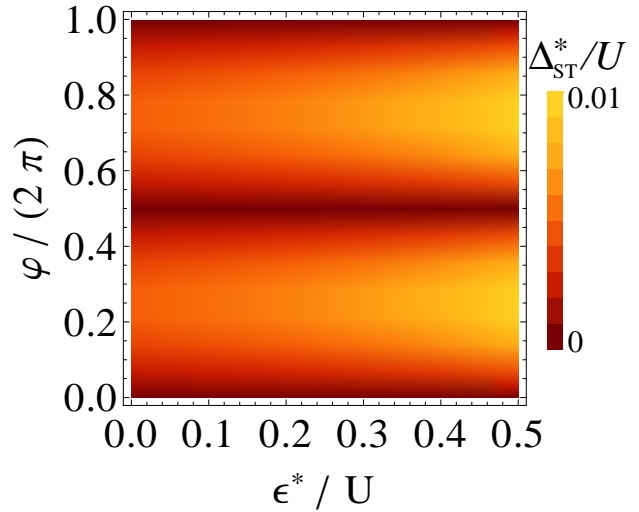}
\caption{\label{fig:deltast3d} Singlet-triplet
  level splitting  $\Delta^*\sub{ST}=\Delta\sub{ST}(\varepsilon^*,
  \bar{b},\varphi)$, Eq.~(\ref{eq:deltast}), for $\delta \vctr{b}=0$
  (no nuclear field), as a function of the detuning
  $\varepsilon^*$ at which the crossing occurs
  (cf. Fig.~\ref{fig:cross02}) and of the angle $\varphi$ between the
  spin-orbit field $\vctr{\Omega}$ and the magnetic sum-field
  $\bar{\bf b}$ (cf. Fig.~\ref{fig:axes}).  Parameters used for this
  plot are $U=1$, $t=0.01$, $V_+=0.75$, $\delta\vctr{b}=0$.}
\end{center}
\end{figure}

Let us first consider the special case of vanishing difference field,
$\delta \vctr{b}=0$, and finite uniform field, $\bar{\vctr{b}} \neq
0$.  In this case, the splitting depends not only on
$\boldsymbol{\Omega}$ but also on the detuning at the minimal
splitting, $\varepsilon^*$, which itself is implicitly determined by
the  sum-field amplitude $\bar{{b}}$.  The geometry of
the system enters through the angle $\varphi$ between $\bar{b}$ and
the spin-orbit field $\boldsymbol{\Omega}$.  In
Fig. \ref{fig:deltast3d} we show a plot  of the level splitting
$\Delta\sub{ST}(\varepsilon^*, \bar{b},\varphi)$, given in
Eq.~(\ref{eq:deltast}) as a function of its variables and for $\delta
\vctr{b}=0$.  

\begin{figure}[t]
\begin{center}
\vskip 0.4cm
\includegraphics[width=8.5cm]{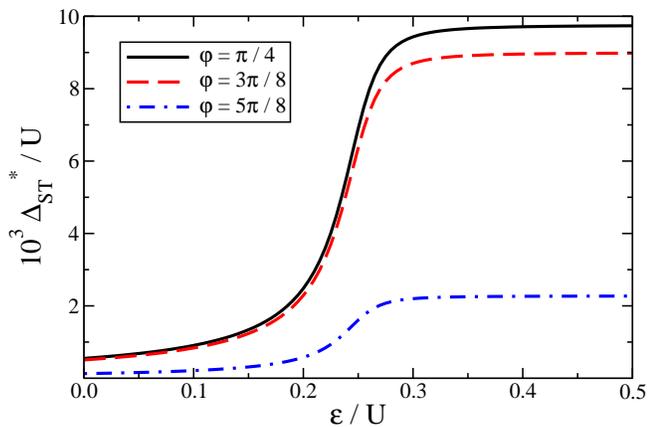}
\caption{\label{fig:nodbbias} Singlet-triplet level
  splitting  $\Delta^*\sub{ST}=\Delta\sub{ST}(\varepsilon^*,
  \bar{b},\varphi)$, Eq.~(\ref{eq:deltast}),  for $\delta \vctr{b}=0$
  (no nuclear polarization), as a function of the detuning
  $\varepsilon^*$ at which the crossing occurs
  (cf. Fig.~\ref{fig:cross02}) and of the angle $\varphi$ between the
  spin-orbit field $\vctr{\Omega}$ and the magnetic sum-field
  $\bar{\bf b}$ (cf. Fig.~\ref{fig:axes}).  At small detuning
  $\varepsilon^* < t$, the splitting becomes rather small, while it
  saturates at large detuning $\varepsilon^*>U$.  The saturation value
  is $\propto |\sin \varphi|$, shown for $\varphi = \pi/4$ (full
  line), $\varphi=3\pi/8$ (dashed line) and $\varphi=5\pi/8$
  (dashed-dotted line).  Parameters used for this plot are $U=1$,
  $t=0.01$, $V_+=0.75$, $\delta\vctr{b}=0$.}
\end{center}
\end{figure}

At any fixed angle $\varphi\neq 0$, $\Delta^*\sub{ST}$ shows a
dependence on the detuning $\varepsilon^*$ at the anticrossing due to
the mixing of $|(0,2)S\rangle$, which is coupled to $|T_+\rangle$
via the spin-orbit interaction, and $|(1,1)S\rangle$, which does not
couple to $|T_+ \rangle$ via spin-orbit coupling in the first order,
see Fig.\ref{fig:nodbbias}.  At large values of detuning
$\varepsilon^*\gg U-V_+$, the splitting reaches a saturation value
$2\Omega \sin\varphi$.  For typical GaAs quantum dots, reaching this
regime requires strong magnetic fields of $B \gg t$.  At lower values
of the detuning $\varepsilon < U-V_+$, the mixing of the singlet
$|(1,1)S\rangle$ becomes significant, and the spin-orbit coupling
value $\Omega$ cannot be read off directly from the splitting.

The maximal splitting $\Delta^*\sub{ST}$ caused by spin-orbit
interaction is $\Delta^*\sub{ST} = 2 \Omega| \sin\varphi|$, for $\psi
= \pi/2$.  From Eq.~(\ref{eq:pres}) and Eq.~(\ref{eq:lambdalambdapm}),
we find that $\Omega$ is set by the material properties (Dresselhaus
($\beta$) and Rashba ($\alpha$) spin orbit interactions) and the
geometry of the dots.  Assuming that the magnetic field is strong
enough to separate the $|T_0\rangle$ and $|T_-\rangle$ states from the
anticrossing, $U \gg \bar{b} \gg t \gg |\vctr{\delta b}| $, the maximal splitting is ($|\sin
\psi | =1$)
\begin{equation}
\label{eq:deltastmaxp}
\Delta^*\sub{ST} = 
\frac{4t}{3}\frac{l}{\Lambda\sub{SO}}|\sin \varphi| , \quad
(\vctr{\delta b} = 0), 
\end{equation}
where $l$ is the interdot distance.  The numeric factor (of order
unity) is non-universal and depends on the specific dot
geometry. Formula (\ref{eq:deltastmaxp}) is one of the main results of
this paper.  It provides a simple but  useful relation between
quantities that can be determined experimentally, such as
$\Delta^*\sub{ST}$, $t$, $l$, and $\varphi$, and a quantity of
interest - the spin orbit length ${\Lambda\sub{SO}}$.  
This relation could allow the strength of spin-orbit coupling to be
measured experimentally\cite{NFT+10,SPJ+11}, though the geometry
and detuning-dependence must be carefully taken into account in order
to obtain an accurate estimate.  

Let us remark here briefly on the special case of zero detuning, i.e.
$\varepsilon=0$, and weak magnetic fields.  In this case, the
splitting is not described by our calculations which require
sufficiently large separation  of the triplets in energy.  However, in
a slightly different system -- a single quantum dot containing two
electrons -- singlet-triplet coupling which is forbidden by
time-reversal symmetry can be generated by applying a magnetic field,
$\Delta\sub{ST}^* \approx (a\sub{B}/\lambda\sub{SO})E\sub{Z}$
\cite{GKL08}.  This case cannot be recovered from our DQD model with
one orbital per site.  Indeed, here we have seen that in weak fields,
${\bar{b}} \ll t$, the coupling of the two states with single
occupation in each well $|(1,1)S\rangle$ and $|T_+\rangle$ due to the
spin-orbit interaction involves doubly occupied states which are
higher in energy due to the on-site repulsion.  On the other hand, for
a pair of electrons in a single quantum dot, the on-site repulsion is
approximately the same for both states, singlet and triplet.  We note
that for DQDs in weak fields, the Zeeman energy $E\sub{Z}$, occurring
in the splitting for a single dot \cite{GKL08}, gets replaced by the
exchange energy $J$ if the mixing of triplets due to
$\delta\vctr{b}\neq 0$ is neglected.

\subsection{Singlet-triplet splitting $\Delta\sub{ST}^*$ for
  $\vctr{\delta b} \neq 0$} 

\begin{figure}[t]
\begin{center}
\includegraphics[width=8.5cm]{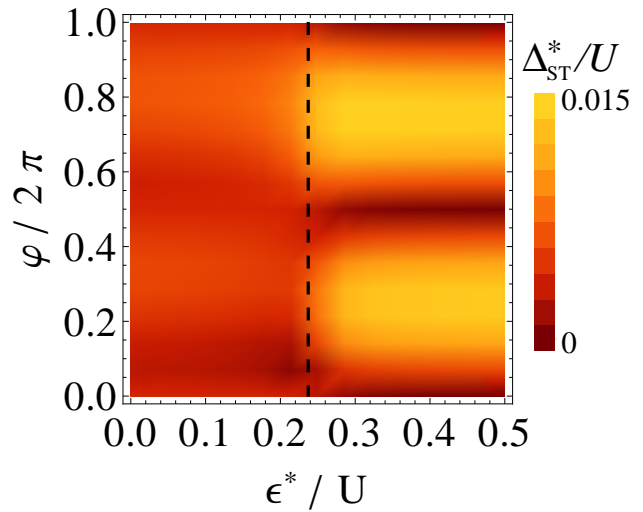}
\caption{\label{fig:deltast1orderfull}
The same plot of the singlet-triplet splitting as in
Fig.~\ref{fig:deltast3d} except for finite nuclear polarization chosen
to be $\delta\vctr{b} = (-0.0006,0.0008,0.0012)$.  In the strong detuning
regime, $\varepsilon^* > U - V_+$ on the right-hand side of the plot, the
splitting is determined by spin orbit interaction, $v\sub{SO} >
v\sub{HF}$ and resembles the
same area in Fig.~\ref{fig:deltast3d}.  In the weak detuning regime,
$\varepsilon^* < U - V_+$ on the left-hand side of the plot, the hyperfine
interaction increases the splitting.  The regime with similar
strengths of the interactions, $v\sub{SO} \sim v\sub{HF}$ can be
identified in the region $\varepsilon^* \approx U-V_+ = 0.25 U$.  For 
$\varepsilon^*$ on the left-hand side of the vertical dashed line, and
$\varphi \approx 0, \pi$, the splitting is dominated by hyperfine
interaction $v\sub{HF} > v\sub{SO}$.  Near the dotted line, and for
the angles $\varphi \approx \pi/2, 3\pi /2$ the sizes of spin orbit
and hyperfine interactions are similar $v\sub{SO}\approx v\sub{HF}$.}
\end{center}
\end{figure}

In addition to spin-orbit coupling of the anticrossing triplet to
$|(0,2)S\rangle$, the anticrossing triplet is coupled to the singlet
$|(1,1)S\rangle$ through the difference-field $\delta\vctr{b}$.  The
previous considerations show that the contributions from the
difference-field $\delta\vctr{b}$ to the splitting cannot be
neglected for angles $\varphi \sim 0,\pi$, or for   field strengths
where $\varepsilon^* < U -V_+$, which is often be the case.
Therefore, we now discuss the splitting in the presence of both, the
spin-orbit field  $\boldsymbol{\Omega}$ and $\delta\vctr{b}$.  The
splitting $\Delta^*_{\rm ST}$ as a function of detuning
$\varepsilon^*$ and direction of $\bar{\vctr{b}}$ is shown in
Fig.~\ref{fig:deltast1orderfull}.  
\begin{figure}[t]
\begin{center}
\includegraphics[width=8.5cm]{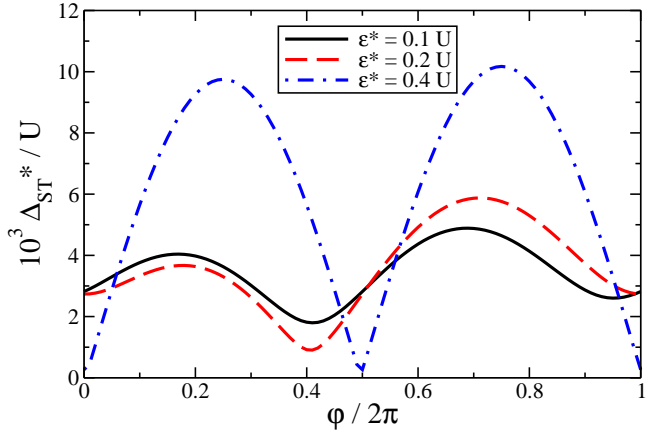}
\caption{\label{fig:asymmetry} First order singlet-triplet splitting
  $\Delta^*\sub{ST}=\Delta\sub{ST}(\varepsilon^*, \bar{b},\varphi)$,
  Eq.~(\ref{eq:deltast}), for finite nuclear polarization
  $\delta\vctr{b}\neq 0$, plotted as  function of the angle $\varphi$
  (cf. Fig.~\ref{fig:axes}), for various detunings $\varepsilon^*$
  (cf. Fig.~\ref{fig:cross02}).  The parameters used for the plot are
  $U=1$, $t=0.01$, $V_+=0.75$, $V_-=0.74$, $\Omega=0.005$, and
  $\delta\vctr{b} = (-0.0006,0.0008,0.0012)$.  The curves correspond to
  $\varepsilon^* = 0.1$ (full line), $\varepsilon^* = 0.2$ (dashed),
  $\varepsilon^* = 0.4$ (dashed-dotted).  In the strong detuning regime
  (dashed-dotted line) the angular dependence reflects the
  $|\sin \varphi|$ dependence of the spin-orbit term
  (see Eq.~(\ref{eq:deltastmaxp}).  The regime of weaker detuning
  (full and dashed line) shows the hyperfine effects.  }
\end{center}
\end{figure}

When both sources of splitting are present, generically, the gap
$\Delta^*\sub{ST}$  remains open.  The spin-orbit contribution to
Eq.~(\ref{eq:deltast}) is always purely imaginary, while the
$\delta\vctr{b}$-contribution has both a real part, coming from the
component lying in the $xz$-plane, and an imaginary part, coming from
the perpendicular component $\delta b_y$.  For any fixed value of the
spin-orbit coupling strength, closing the gap would require fine
tuning of $\delta \vctr{b}$, both in amplitude and
direction.  As a function of the direction of $\bar{\vctr{b}}$ the
contributions to $\Delta^*\sub{ST}$  compete, and, in addition, the
relative size of the competing terms will change as a function of
$\varepsilon^*$.  Indeed, the spin-orbit term is strongest at  large
detuning $\varepsilon^* \gg U-V_+$, while the difference-field term
becomes significant at small detuning $\varepsilon^*\sim t$.  This
competition  affects the form of the $\varphi$-dependent splitting,
see Fig.\ref{fig:asymmetry}.

\begin{figure}[t]
\begin{center}
\includegraphics[width=8.5cm]{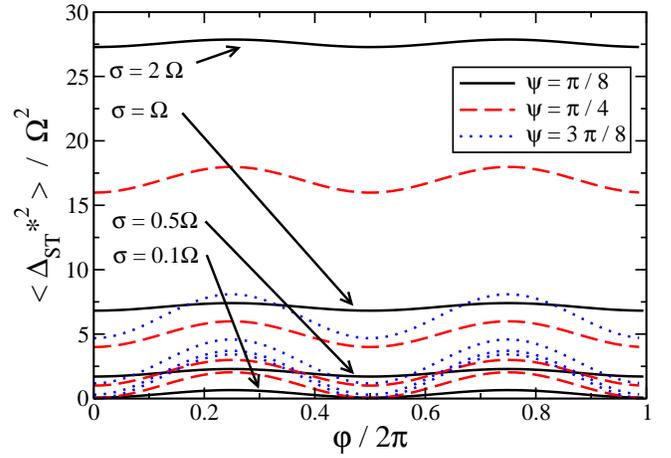}
\caption{\label{fig:averageSquaredPhi} Square of the splitting $\langle
  \Delta\sub{ST}^{*2} \rangle$, Eq.~(\ref{eq:deltast}), averaged over
  Gaussian fluctuations of $\delta\vctr{b}$ with zero mean and
  standard deviation $\sigma$.  The plots show the dependence of
  $\langle \Delta\sub{ST}^{*2} \rangle$ on the angle $\varphi$ for
  various mixing angles $\psi$.  We have assumed isotropic Gaussian
  fluctuations with  a standard deviation $\sigma$.  Plots are for the
  values $\sigma/\Omega = 0.1, 0.5, 1, 2$ and illustrate the effects
  of various strengths of fluctuations.  The curves are found by
  numerical averaging  over the fields $\delta\vctr{b}$.}
\end{center}
\end{figure}

In the limit $|\vctr{\Omega}\sin \psi| \ll |\delta \vctr{b} \cos
\psi|$, the splitting $\Delta^*\sub{ST}$ is caused mostly by the inhomogeneous
field.  In this case, the splitting is proportional to the size of component
$\delta\vctr{b}\sub{\perp}=\delta\vctr{b} -
\vctr{e}(\delta\vctr{b}\cdot \vctr{e})$ of $\delta\vctr{b}$ which is
normal to the homogeneous field $\bar{\vctr{b}}=|\bar{\vctr{b}}| \vctr{e}$.
With the leading spin-orbit coupling correction, the splitting is [see
  Eq.(\ref{eq:deltast})]  
\begin{equation}
\label{eq:dstlowestomega}
\Delta^*\sub{ST} = 2\sqrt{2}|\delta\vctr{b}_{\perp} \cos{\psi} | - \frac{ 2 \Omega
  \delta b_y \sin\psi \cos\psi}{|\delta \vctr{b}_{\perp} \cos\psi| }.
\end{equation}

\subsection{Measuring the singlet-triplet coupling}

The singlet-triplet coupling $\Delta\sub{ST}^*$ is manifested
experimentally, for example, in the spin flip probability when the
system is taken through the level crossing during a time-dependent
gate sweep \cite{PTJ+08}.  In such experiments, the system is
initialized to its ground state at large $\epsilon$, the $(0,2)$
singlet.  When $\epsilon$ is then ramped to take the system through
the singlet-triplet crossing, the two-electron spin state may change,
with a probability determined by a combination of the coupling
$\Delta\sub{ST}^*$ and the sweep rate.  The final spin state can then
be read out by quickly ramping back to large $\epsilon$, where the
singlet and triplet states have discernibly different charge
distributions, which can be detected by a nearby charge sensor.  

Even with single-shot spin detection \cite{BRM+09}, determining the
spin flip probability requires building up statistics over many
experimental runs.  Within each run, the parameters in
Eq.~(\ref{eq:deltast}) may be considered fixed.  However, the
hyperfine field components are in
general {\it a priori} unknown: under typical experimental conditions,
the temperature is high compared with all intrinsic energy scales
within the nuclear spin system, and the equilibrium state is nearly
completely random.  Depending on the measurement timescale, the
nuclear fields on subsequent experimental runs may either remain
constant or  may change.  While the correlation time for the
longitudinal component of the nuclear field (parallel to the external
field) may be quite long, the transverse components change on the
timescale of nuclear Larmor precession, which for moderate fields of a
few hundred millitesla can reach the sub-microsecond timescale.  The
coherence time associated with this precession may reach several
hundred microseconds to one millisecond. 

Let $P(\Delta^*_{\rm ST})$ be the  probability that the system makes a
transition to the triplet state in a single sweep, when the value of
$\Delta^*_{\rm ST}$ is specified. In an experiment where measurements of the
singlet and triplet fractions are averaged over a time long compared to
all nuclear spin relaxation times, one obtains an averaged  probability
$\langle P(\Delta^*_{\rm ST})\rangle$, where $\langle A\rangle$
denotes  the mean value of  quantity $A$, averaging over a Gaussian
distribution of  $\delta \vctr{b}$, while other parameters such as $B,
t, \bar{b}, \phi$ and the sweep rate  are held fixed.  If the
measurements are averaged over a shorter period, which is long
compared to the time for phase relaxation of the nuclear spins, but
short compared to the longitudinal relaxation times, then the Gaussian
average should be taken   only over the transverse components of
$\delta \vctr{b}$, while the component parallel to the applied
magnetic field is held fixed.  

When the sweep rate through the S-T transition is rapid, the probability
$P(\Delta^*_{\rm ST})$ should be proportional to $(\Delta^*_{\rm ST})^2$, so an
average value of $P(\Delta^*_{\rm ST})$ will measure the mean value of
$(\Delta^*_{\rm ST})^2$.  For lower values of the sweep rate, $P$ will have
corrections due to  $(\Delta^*_{\rm ST})^4$, etc..  Therefore, measurements of
the averaged value $\langle P(\Delta^*_{\rm ST}) \rangle$  for a wide
range of sweep rates should, in principle, yield average values of all
powers of $(\Delta^*_{\rm ST})^2$, and thus allow one to deduce the
probability distribution for $(\Delta^*_{\rm ST})^2$.  Here we
concentrate on the mean value of $(\Delta^*_{\rm ST})^2$ , and discuss
predictions for this mean value as a function of the parameters $B,
\phi,$ and $t$.  

\begin{figure}[t]
\begin{center}
\includegraphics[width=8.5cm]{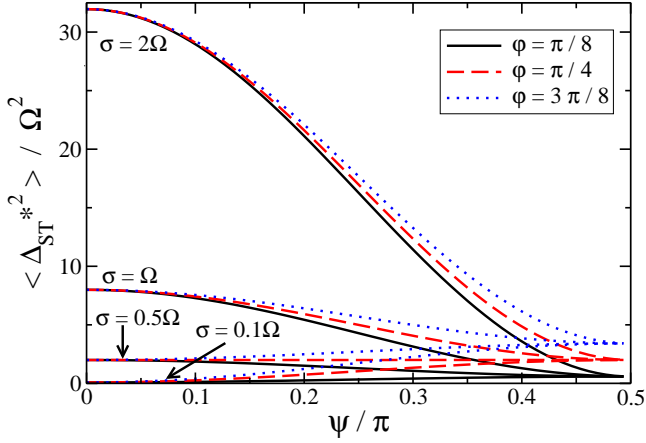}
\caption{\label{fig:averageSquaredPsi} Average square of the splitting
  $\langle\Delta\sub{ST}^{*2} \rangle$, in the fluctuating nuclear field
  $\delta\vctr{b}$.  The parameters are chosen as in
  Fig.~\ref{fig:averageSquaredPhi}, and we illustrate the dependence on the
  mixing angle $\psi$.}
\end{center}
\end{figure}

We illustrate the dependence of $\langle (\Delta\sub{ST}^*)^2 \rangle$
on the angle $\varphi$ and mixing $\psi$ in
Fig.~\ref{fig:averageSquaredPhi} and Fig.~\ref{fig:averageSquaredPsi}.
The dependence of the splitting on the angle $\varphi$, inherent in
the non-averaged splitting, see Eq.~(\ref{eq:deltast}), remains
visible when the splitting is dominated by spin-orbit interaction.  As
expected, the dependence of the splitting on the angle $\varphi$,
Fig.~\ref{fig:averageSquaredPhi}, is most visible in the case of weak
fluctuations of $\vctr{\delta b}$, i.e. for weak hyperfine coupling.
In addition, the angular dependence is more pronounced for larger
mixing angles, since the spin-orbit induced splitting depends on
$\Omega\sin\psi$, Fig.~\ref{fig:averageSquaredPsi}.   

The fluctuating difference field, besides changing the average $\langle
(\Delta\sub{ST}^*)^2 \rangle$, introduces noise in the splitting.  We find
that the standard deviation $\sigma_{\Delta}$ of the splitting, in the
limit of weak fluctuations $|\Omega\sin\varphi\sin\psi|\gg \sigma$ is
\begin{equation}
\label{eq:deltastspreadgauss}
\sigma\sub{\Delta,G} = 2 \sigma |\cos \psi|,
\end{equation}
so that it also shows dependence on the mixing angle $\psi$.

Fluctuations in $\delta\vctr{b}$ smear the splitting at the
anticrossing $\Delta\sub{ST}^*$.  The average value and noise in the
splitting can be used to measure the strengths of spin-orbit coupling
$\Omega$ and the hyperfine field $\delta\vctr{b}$.  The relative size
of fluctuations in $\Delta\sub{ST}^*$, as a function of $\varphi$, has
minima at $\varphi \approx \pi/2, 3\pi/2$. 

\subsection{Higher order corrections}

So far, our analysis of the $|S_-\rangle$ - $|T_+\rangle$ anticrossing
was based on the assumption that the largest contribution to the
splitting $\Delta^*\sub{ST}$, Eq.~(\ref{eq:deltast}), results from the
direct coupling of the two states via $H\sub{ST}\subup{SO}$ and
$H\sub{ST}\subup{\delta\vctr{b}}$.  However, if the detuning
$\varepsilon ^*$ is not large enough to make the influence of the
levels that are energetically further away  from the anticrossing
completely negligible,  higher order terms  that describe virtual transitions to
such higher levels and thus involve more than one transition between
the singlets and the triplets become important.

To study this regime, we derive an effective Hamiltonian in the
vicinity of the anticrossing by a second order Schrieffer-Wolff (SW)
transformation\cite{SW66,BDL11}.  We divide the Hilbert space of the
DQD into a relevant part which includes  the anticrossing states, and
an auxiliary part which contains the remaining 4 states.  The
time-independent perturbation series is then performed in powers of
the spin-non-conserving interactions.  The spin-conserving Hamiltonian
$H\sub{sc}$, Eq.~(\ref{eq:h1120spinconservingblocks}), is taken as the
unperturbed part, while $H-H\sub{sc}$ is the perturbation.  In
reordering the basis, we choose the crossing states $|S_-\rangle$ and
$|T_+\rangle$ of $H\sub{sc}$ to be the first two basis states.  Then,
the Hamiltonian has a block-diagonal form denoted by
\begin{equation}
\label{eq:hforswblocks}
H= \left(
\begin{array}{cc}
A & C
\\
C ^{\dagger} & B
\end{array}
\right),
\end{equation}
where $A$ is a $2\times 2$ matrix that describes the anticrossing
states, $B$ is a $4\times 4$ matrix of the states with energies far
from the anticrossing and the $2\times 4$ matrix $C$ represents the
coupling between of the subspaces controlled by $A$ and $B$.  In the
basis $\left( |S_-\rangle, |T_+\rangle, |T_0\rangle, |T_-\rangle,
|(2,0)S\rangle, |S_+\rangle \right)$, with the two anticrossing level
at the positions $1$ and $2$, we can read off the blocks from
Eq.~(\ref{eq:hforswblocks}): 
\begin{widetext}
\begin{equation}
\label{eq:ablock}
A= \left(
\begin{array}{cc}
E_{S-}
&
-i\Omega_s \sin \psi + \sqrt{2} \delta_+ \cos \psi
\\
i\Omega_s \sin \psi + \sqrt{2} \delta_- \cos \psi
&
E_{T+}
\end{array}
\right),
\end{equation}
\begin{equation}
\label{eq:bblock}
B= \left(
\begin{array}{cccc}
V_-
&
0
&
i \sqrt{2} \Omega_c
&
-i \sqrt{2} \Omega_c \cos \psi + 2 \delta \sin \psi
\\
0
&
V_- + \bar{b}
&
-i \Omega_s
&
i \Omega_s \cos \psi - \sqrt{2} \delta_+ \sin \psi
\\
-i \sqrt{2} \Omega_c
&
i \Omega_s
&
U + \varepsilon
&
- \sqrt{2} t \sin \psi - X \cos \psi
\\
i \sqrt{2} \Omega_c \cos \psi + 2 \delta \sin \psi 
&
- i \Omega_s \cos \psi - \sqrt{2} \delta_- \sin \psi
&
-\sqrt{2} t \sin \psi - X \cos \psi
&
E_{S+}
\end{array}
\right),
\end{equation}
\begin{equation}
\label{eq:cblock}
C= \left(
\begin{array}{cccc}
-i \sqrt{2} \Omega_c \sin \psi + 2 \delta \cos \psi
&
i \Omega_s \sin \psi - \sqrt{2} \delta_- \cos \psi
&
-\sqrt{2} t \cos \psi + X \sin \psi
&
0
\\
0
&
0
&
i \Omega_s
&
-i \Omega_s \cos \psi + \sqrt{2} \delta_- \sin \psi
\end{array}
\right),
\end{equation}
\end{widetext}
where we have used the
abbreviations $\delta= \delta \vctr{b} \cdot \vctr{e}$, $\delta_{\pm}=
\delta \vctr{b} \cdot \vctr{e'} \pm i \delta b_y$, $\Omega_s = \Omega
\sin \varphi$, $\Omega_c = \Omega \cos \varphi$, and the unit vectors
$\vctr{e'}$ and $\vctr{e}$ are defined in Eq.~(\ref{eq:eprimedef}) and
Eq.~(\ref{eq:edef}), respectively.

As a result of the  SW transformation on Eq.~(\ref{eq:hforswblocks}),
the off-diagonal block $C$ is eliminated up to second order in $C$ and
the transformed block $A\xrightarrow{\rm SW} A + \delta A$ becomes
then the  Hamiltonian of an effective two-level system.  Therefore,
the first-order Hamiltonian $H\sub{cr}^{(1)}$ from Eq.~(\ref{eq:hcr1})
becomes modified by second order terms, $H\sub{cr} \xrightarrow{\rm
  SW} H\sub{cr} + \delta A$, where $\delta A$ is the second order
correction to $A$. The diagonal matrix elements $\delta A_{11}$ and
$\delta A_{22}$ describe the renormalization  of the energy levels,
and their effect is to shift the detuning $\varepsilon^*$ at which the
anticrossing occurs.  The explicit expressions of these corrections
are
\begin{widetext}
\begin{align}
\begin{split}
\label{eq:deltaa11}
\delta A_{11} =& \frac{1}{E_{S-} - V_-} \left[ 2 \left( \Omega_c \sin \psi \right)^2 + 4 \delta ^2 \cos^2 \psi \right] + \frac{1}{E_{S-} - U - \varepsilon} \left( \sqrt{2} t \cos \psi - X \sin\psi \right)^2 +
\\
&+
\frac{1}{E_{S-} - V_- - \bar{b} } \left[ 2 \left( \delta \vctr{b} \cdot \vctr{e'} \right)^2 \cos^2 \psi + \left( \sqrt{2} \delta b_y \cos \psi + \Omega_s  \sin \psi \right)^2 \right],
\end{split}
\\
\label{eq:deltaa22}
\delta A_{22} =& \frac{1}{E_{T_+} - U - \varepsilon } \Omega_s^2 + \frac{1}{E_{T+} - E_{S+}} \left[ \left( \Omega_s \cos \psi + \sqrt{2} \delta b_y \sin \psi \right)^2 + 2 \left( \delta \vctr{b} \cdot \vctr{e'} \right) ^2 \sin^2 \psi \right],
\\
\label{eq:deltaa12}
\delta A_{12} =& -\frac{i}{2} \Omega_s \left( \sqrt{2} t \cos \psi + X \sin \psi \right) \left( \frac{1}{E_{S-} - U - \varepsilon} + \frac{1}{E_{T+} - U - \varepsilon} \right). 
\end{align}
\end{widetext}

\begin{figure}[t]
\begin{center}
\includegraphics[width=8.5cm]{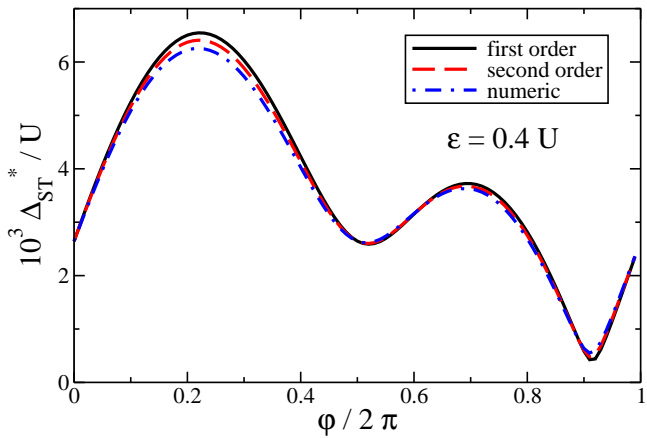}
\caption{\label{fig:12e} Comparison of the splitting $\Delta\sub{ST}^*$
  obtained from the perturbation in first order (full line),
  Eq.~(\ref{eq:deltast}), and in the second order (dashed line),
  Eq.~(\ref{eq:deltast2}) to the exact numerical result (dashed-dotted
  line), obtained by the direct numerical diagonalization of $H$,
  Eq.~(\ref{eq:hallmatrixoriginal}).  The plots show the splitting
  $\Delta\sub{ST}^*$ as a function of the angle $\varphi$ for the
  anticrossing at the detuning $\epsilon^*=0.4 U$.  The parameters
  used in this plot are $U=1$, $t=0.02$, $V_+=0.75$, $V_-=0.74$,
  $\Omega=0.005$, and $\delta\vctr{b} = (-0.0006,0.0008,0.0012)$.}
\end{center}
\end{figure}

In experiments that probe the electron spin dynamics, the most
important terms are the off-diagonal ones, $\delta A_{12} = \delta
A_{21}^{*}$.  They lead to a modification of the first order
singlet-triplet splitting Eq.~(\ref{eq:deltastar}),
i.e. $\Delta^*\sub{ST} \xrightarrow{\rm SW} 2|H_{34} + \delta
A_{12}|$.  Thus, up to second order, the splitting at the anticrossing
becomes
\begin{widetext}
\begin{equation}
\label{eq:deltast2}
\Delta \sub{ST}^* = 2\left| - i\Omega \sin \varphi  \left[ \sin \psi -
  \left(  \frac{t}{\sqrt{2}} \cos \psi - \frac{X}{2} \sin \psi \right)
  \left( \frac{1}{E_{S-} - U - \varepsilon^*} + \frac{1}{E_{T+} - U -
    \varepsilon^*} \right) \right] + \sqrt{2} \left( \delta \vctr{b}
\cdot \vctr{e'} + i \delta b_y \right) \cos \psi \right|.
\end{equation}
\end{widetext}
We see now that the new correction terms in $\Delta \sub{ST}^*$ become
significant for weak detuning, when $\psi < \pi/2$, because the
spin-independent tunneling contribution, which is $\propto \cos\psi $,
can alter the first-order result, which is $\propto \sin \psi$.  We
compare the splitting in the second order, Eq.~(\ref{eq:deltast2}),
with the first order splitting and the result of exact numerical
diagonalization of $H$, Eq.~(\ref{eq:hallmatrixoriginal}), in
Fig.~\ref{fig:12e}.  The higher order corrections are small, but they
do become significant for the magnetic field normal to the spin orbit
parameter, $\varphi =\pi/2,3\pi/2$, due to stronger effective strength $\Omega
\sin\varphi$ of spin-orbit coupling.  For other considered values of
detuning, $\varepsilon^* = 0.1 U$ and $\varepsilon^* =  0.2 U$, the
change of splitting is smaller than for the $\varepsilon^* = 0.4 U$
case.  

The limit $\psi = \pi/2$ requires strong magnetic fields,
${B}\sim 1\unit{T}$ for a typical GaAs DQD.  It is reasonable to assume
that the experiments can be performed both in this limit and away from it, so
that the dependence of $\Delta\sub{ST}^*$ on $\psi$ can be probed.  In
materials with larger g-factors such as InAs, InSb, SiGe, the limit is
reached at lower fields.  In addition, we have obtained similar
results for a model DQD with $t = 0.1 U$, $\Omega = 0.1 t$,
$|\delta\vctr{b}| \approx 0.5 \Omega$, that describes a smaller DQD
with more pronounced hopping.

\section{Ratio of spin-orbit and hyperfine terms}
\label{sec:switching}

As pointed out before, the  spin non-conserving Hamiltonian in the
vicinity of the anticrossing can be used to get experimental access to
the spin orbit interaction and nuclear polarization in the
difference-field $\delta{\vctr{b}}$.  Being a function of controllable
parameters $\epsilon$ and $\bar{\vctr{b}}$, this Hamiltonian can be
altered by applying voltages to electrodes in the vicinity of the
quantum dots, adjusting the strength of an external magnetic field, or
changing the direction of the field.  

The effect of the competition between spin orbit and hyperfine induced
spin flips on the efficiency of angular momentum transfer between
electron and nuclear spins  was recently studied theoretically, both
in the context of dc transport experiments \cite{RL10} and in the
context of gate-sweep experiments \cite{RNL+10,BR11}.   References
\cite{RL10} and \cite{RNL+10} revealed striking sensitivities of the
polarization transfer efficiency on the ratio of spin orbit and
hyperfine coupling strengths.  In those works, the coupling strengths
were treated as phenomenological parameters.    Here we provide
explicit expressions for them, and discuss how they can be tuned.

Following the notation of Ref.~\cite{RNL+10} we write
\begin{equation}
\label{eq:vstdef}
v_{\vartheta} = v\sub{SO} + e^{i\vartheta} v\sub{HF},
\end{equation}
where $v\sub{SO}$ and $v\sub{HF}$ stand for the transitions caused by
spin orbit and hyperfine interactions, respectively.  Comparing with
Eq.~(\ref{eq:hcr1}), we can identify $v_{\vartheta}$ with $H\sub{34}$.
Then, adjusting the overall phase to make the spin-orbit part
$v\sub{SO}$ real, we identify in lowest order
\begin{align}
\label{eq:vso}
v\sub{SO} &= |\Omega \sin \varphi \sin \psi | \\
\label{eq:vhf}
v\sub{HF} &= |\cos \psi| \sqrt{\left( \delta \vctr{b} \cdot \vctr{e'}
  \right)^2 + \delta b_y^2 }, 
\\
\label{eq:vartheta}
\vartheta &= \arctan \left[ \frac{-\delta \vctr{b} \cdot \vctr{e'}}{\delta b_y} \right]. 
\end{align}
The explicit expression for $\vartheta$ shows that the phase of the
matrix element can be adjusted not only by changing the direction of
$\delta \vctr{b}$, but also by rotating the external magnetic field,
which changes $\vctr{e'}$, and also controls the effective spin-orbit
coupling strength.

Our results show that, in principle,  it is possible to switch between
the two regimes of hyperfine-dominated and spin-orbit-dominated
behavior, by changing the external magnetic field strength and
direction.  In the strong  field regime, with sum-field
$\bar{\vctr{b}}$ being large, $\varepsilon^*$ is also large, and thus
the spin-orbit terms become dominant ($\psi$ approaches $\pi/2$).
This behavior is illustrated in Fig.~\ref{fig:deltast1orderfull}.
Note that the large values of $\varepsilon^*$ require strong $\bar{b}$
fields.  On the other hand, switching to the
regime $v\sub{HF} > v\sub{SO}$ is always possible by rotating the
direction of the magnetic field so that it coincides with $\pm
\boldsymbol{\Omega}/|\boldsymbol{\Omega}|$ giving $\varphi\approx 0$.
In this case, the term $v\sub{SO}$ is negligible and $v\sub{HF}$
dominates the splitting.  Higher-order corrections to the effective
Hamiltonian at the anticrossing point do not alter this basic picture
of the splitting, but they do change the values of the parameters
$\psi$ and $\varphi$ at which the switching occurs. 

The switching between the regimes dominated either by spin orbit or by
hyperfine interactions can potentially be achieved as follows.  For the  regime
$v\sub{SO}> v\sub{HF}$, the sum-field $\bar{\vctr{b}}$ should point
along the spin orbit field $\boldsymbol{\Omega}$, see
Eqs.~(\ref{eq:omegadef}) and (\ref{eq:etadefsbd+}), in order to
maximize $|\sin \varphi|$.  Also, the applied field should be as
strong as possible, in order to maximize the amplitude of the singlet
$|(0, 2)S\rangle$ (contributing to the anticrossing  singlet
$|S_-\rangle$, see Eq.~(\ref{eq:spmdef})).  On the other hand, the
opposite regime, $v\sub{HF}>v\sub{SO}$, can be reached by orienting
$\bar{\vctr{b}}$ along $\boldsymbol{\Omega}$, and thus reaching $\sin
\varphi =0$.  If $\sin \varphi =0$ cannot be achieved, $v\sub{SO}$
can be reduced by decreasing  $\bar{\vctr{b}}$  and thereby increasing
the amplitude of the singlet state $|(1,1)S\rangle$ in the
$|S_-\rangle$-singlet at the anticrossing.

\section{Conclusions}
\label{sec:conclusions}

We have derived an effective two level Hamiltonian $H\sub{cr}$ for a
detuned two-electron double quantum dot in an external magnetic field.
Our effective Hamiltonian describes the dynamics of the electron spins
for the values of detuning $\varepsilon\approx \varepsilon^*$ close to
the anticrossing of the lowest energy $S=0$ and the lowest energy
$S=1$ state.  We have shown how $H\sub{cr}$ can be used in the
interpretation of experiments that probe electron spin interactions by
charge sensing and transport in the Coulomb blockade regime.  The
dependence of $H\sub{cr}$ on the detuning and magnetic fields can also
be used to switch the spin dynamics in a double quantum dot between
the spin-orbit dominated regime, and the hyperfine-dominated regime.

The spin dynamics at the anticrossing is governed by the spin-orbit
and nuclear hyperfine interactions.  In a double quantum dot, these
two interactions act differently on the orbital electronic states.  On
one hand, the spin-orbit interaction causes hopping of an electron
between the quantum dots accompanied by a spin rotation, thus changing
the occupation of the quantum dots.  On the other hand, the nuclear
hyperfine interaction acts as an inhomogeneous magnetic field, and
causes spin rotations that are local to the dots, leaving the charge
state unchanged.  Due to this distinction, the detuning $\varepsilon$
controls the relative strength of the two interactions in $H\sub{cr}$,
in addition to the ratio $|\Omega|/|\delta\vctr{b}|$, or
$|\Omega|/\sigma$.  In the limit of detuning much stronger than the
on-site repulsion of the dots, $\varepsilon \gg U$, $H\sub{cr}$
describes mostly the spin-orbit interaction, with negligible hyperfine
effects.  In the case of weaker detuning, the effective hyperfine
interactions can be of the size comparable to the effective spin orbit
interactions.  

In addition, we find that the orientations of both the double quantum
dot and the external magnetic field, described in the $H\sub{cr}$ by
the spin-orbit field $\boldsymbol{\Omega}$ and the sum field
$\bar{\vctr{b}}$, influence the effective spin orbit interaction.  In
particular, by having  $\bar{\vctr{b}}$ pointing along
$\boldsymbol{\Omega}$, we can suppress the spin orbit  effects
completely (in leading order).

The splitting of the anticrossing states is accessible to experiments.
It can be calculated from $H\sub{cr}$, and we find the dependence of
this splitting on detuning and the strength and direction of the
sum field, $\Delta\sub{ST}(\varepsilon,\bar{b},\varphi)$.
Of particular interest is the splitting of levels at the
anticrossing.  We calculate this quantity,
$\Delta^*\sub{ST}(\varepsilon^*,\varphi)$, as a function of the
detuning at the anticrossing point, $\varepsilon^*$, and the
orientation of the sum field, given by the angle $\varphi$.  Both the
spin orbit interaction strength and the inhomogeneity in the hyperfine
coupling can be deduced by measuring the splitting and using our
formulas for $\Delta^*\sub{ST}(\varepsilon^*,\varphi)$.

The relative strength of the spin orbit and hyperfine terms in
$H\sub{cr}$ has a profound effect on the coupled dynamics of electron
and nuclear spins.  The value of the average angular
momentum transfer to nuclear spins as an electron tunnels through a
spin-blockaded DQD changes sharply as the interaction goes from the
spin-orbit-dominated to the hyperfine-dominated regime.  The spin
orbit interaction is dominant in the limit of strong detuning
$\varepsilon^*\gg U -V_+$.  The regime dominated by nuclear hyperfine
interaction is reached when the detuning is weaker $\varepsilon^*
\lesssim t$ and the orientation of the sum field is along
$\boldsymbol{\Omega}$.  Using the dependencies of the matrix elements
on gate voltages and magnetic field strength and orientation, it may
be possible to tune between these two regimes {\it in situ}, thus
enabling experiments to study their sensitive competition.

{\it Acknowledgements}  We gratefully acknowledge helpful discussions
with Izhar Neder.  This work is partially supported by the Swiss NSF, NCCR Nanoscience and QSIT, DARPA QuEST, and the Intelligence Advanced Research Projects Activity (IARPA) through the Army Research Office.

\bibliography{all}

\bibliographystyle{apsrev}

\end{document}